\documentclass[runningheads]{llncs}

\usepackage{eccv}

\usepackage{eccvabbrv}

\usepackage{graphicx}
\usepackage{booktabs}

\usepackage[accsupp]{axessibility}  %

\usepackage{hyperref}

\usepackage{orcidlink}

\newif\ifarxivsubmit
\arxivsubmitfalse

\newif\ifdraft
\draftfalse

\newif\ifsuppsubmit
\suppsubmittrue

\newif\ifsigsubmit
\sigsubmitfalse

\ifsigsubmit
    \def\fakepara{\paragraph}

    \usepackage[capitalize]{cleveref}
    \usepackage{xspace}
    
    \makeatletter
    \DeclareRobustCommand\onedot{\futurelet\@let@token\@onedot}
    \def\@onedot{\ifx\@let@token.\else.\null\fi\xspace}
    
    \def\eg{\emph{e.g}\onedot} 
    \def\ie{\emph{i.e}\onedot}

    \def\etal{\emph{et al}\onedot}
    \makeatother

\else
    \newcommand{\fakepara}[1]{\smallskip\noindent\textbf{#1}}
    \ifarxivsubmit
    \else
        \usepackage[accsupp]{axessibility}
    \fi
\fi

\ifdraft
    \newcommand{\namanh}[1]{{\color{orange}[\textbf{Nam Anh:} #1]}}
    \newcommand{\itai}[1]{{\color{magenta}[\textbf{Itai:} #1]}}
    \newcommand{\oded}[1]{{\color{purple}[\textbf{Oded:} #1]}}
    \newcommand{\rana}[1]{{\color{cyan}[\textbf{Rana:} #1]}}

    \newcommand{\rh}[1]{{\color{cyan}#1}}
\else
    \newcommand{\namanh}[1]{}
    \newcommand{\itai}[1]{}
    \newcommand{\oded}[1]{}
    \newcommand{\rana}[1]{}

    \newcommand{\rh}[1]{{#1}}
\fi

\newcommand{\ourmethod}{RADmesh}
\newcommand{\ourtitle}{\ourmethod: Remesh-Aware Mesh Deformation}

\ifsigsubmit
    \let\titleold\title
    \renewcommand{\title}[1]{\titleold{#1}\newcommand{\thetitle}{#1}}
    
\fi

\AtEndPreamble{
    \usepackage[capitalize]{cleveref}
    \crefname{section}{Sec.}{Secs.}
    \Crefname{section}{Section}{Sections}
    \Crefname{table}{Table}{Tables}
    \crefname{table}{Tab.}{Tabs.}
}

\RequirePackage{enumitem}
\setlist[itemize]{noitemsep,leftmargin=*,topsep=0em}
\setlist[enumerate]{noitemsep,leftmargin=*,topsep=0em}

\RequirePackage{multirow}
\RequirePackage{multicol}

\newcommand{\caltil}[1]{\ensuremath{\mathcal{{#1}}}} %

\DeclareMathOperator{\SO}{SO}
\DeclareMathOperator*{\argmin}{argmin}

\def\MM{\mathcal{M}}
\def\VV{\mathcal{V}}

\def\NN{\mathcal{N}}
\def\QQ{\mathcal{Q}}
\def\TT{\mathcal{T}}

\def\uu{\mathsf u}
\def\qq{\mathsf q}
\def\pp{\mathsf p}

\def\ee{\mathsf e}

\def\Rr{\mathsf R}
\def\Ss{\mathsf S}
\def\Tt{\mathsf T}

\usepackage{wrapfig}
\renewcommand{\fakepara}[1]{{\smallskip\noindent\textit{#1}}}

\begin{document}

\title{\ourtitle}

\author{
Nam Anh Dinh\inst{1}\and
Itai Lang\inst{1}\and
Oded Stein\inst{2,3}\and
Rana Hanocka\inst{1}
}

\authorrunning{Dinh et al.}

\institute{University of Chicago, Chicago IL 60637, USA\\
\email{\{namanh,itailang,ranahanocka\}@uchicago.edu}
\and
University of Southern California, Los Angeles CA 90089, USA\\
\email{ostein@usc.edu}
\and
Technion, Haifa, Israel\\
\email{oded.stein@technion.ac.il}
}

\maketitle

\begin{center}
    \includegraphics[width=0.99\linewidth]{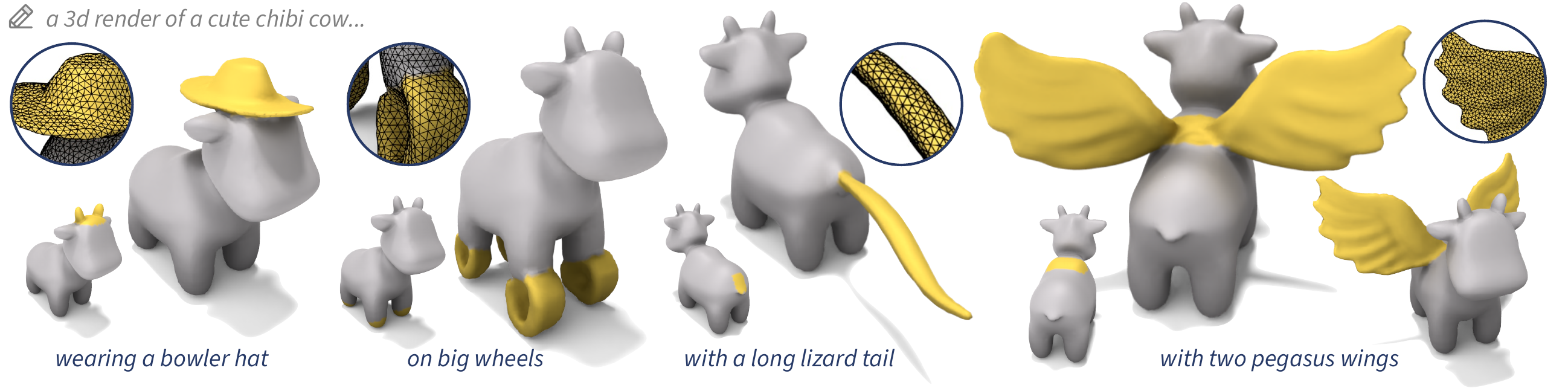}
    \captionof{figure}{By deforming together with remeshing in the loop, our method is capable of growing large, detailed parts and imparting drastic geometric changes on small selected regions (yellow) on the source shape (bottom left of each example), guided by text prompts. The insets show that the growths maintain isotropic triangulations with good element quality, indicating their suitability for the new geometry.}
    \label{fig:teaser}
\end{center}

\begin{abstract}
We propose a remeshing-enhanced method for generatively deforming shapes with visual losses. It is intuitive that sufficiently drastic deformations of a mesh without changing its triangulation can easily compromise element quality, even if such large geometry changes may be semantically desired. Shape deformation methods could thus benefit from changing the triangulation; however, this is not done by most generative, text-based, visually-supervised mesh deformation methods. Remeshing is a discrete operation, proven to be especially challenging to couple with the notoriously noisy supervision signal provided by visual losses. We propose a vertex-based deformation optimization quantity capable of large deformations and robustness to such noise; we periodically remesh using an isotropic remesher that interpolates and carries forward the deformation optimization state. This enables continuous, geometry-informed progress in coarse-to-fine addition of resolution. The resulting shapes' triangulations fit their optimized geometry and have neat isotropic elements. Further, our method is localizable, able to grow new features on a base shape with expressive detail, leaving the rest unchanged. We showcase the effectiveness of our method on a variety of shapes and prompts, both local and global deformations, and demonstrate its superior visual quality and triangle efficiency. Our project page is at \url{https://threedle.github.io/radmesh}.
\end{abstract}

\def\citet{\cite}

\section{Introduction} \label{sec:introduction}

\begin{figure*}
    \centering
    \includegraphics[width=0.99\linewidth,trim=0 20 0 0 clip]{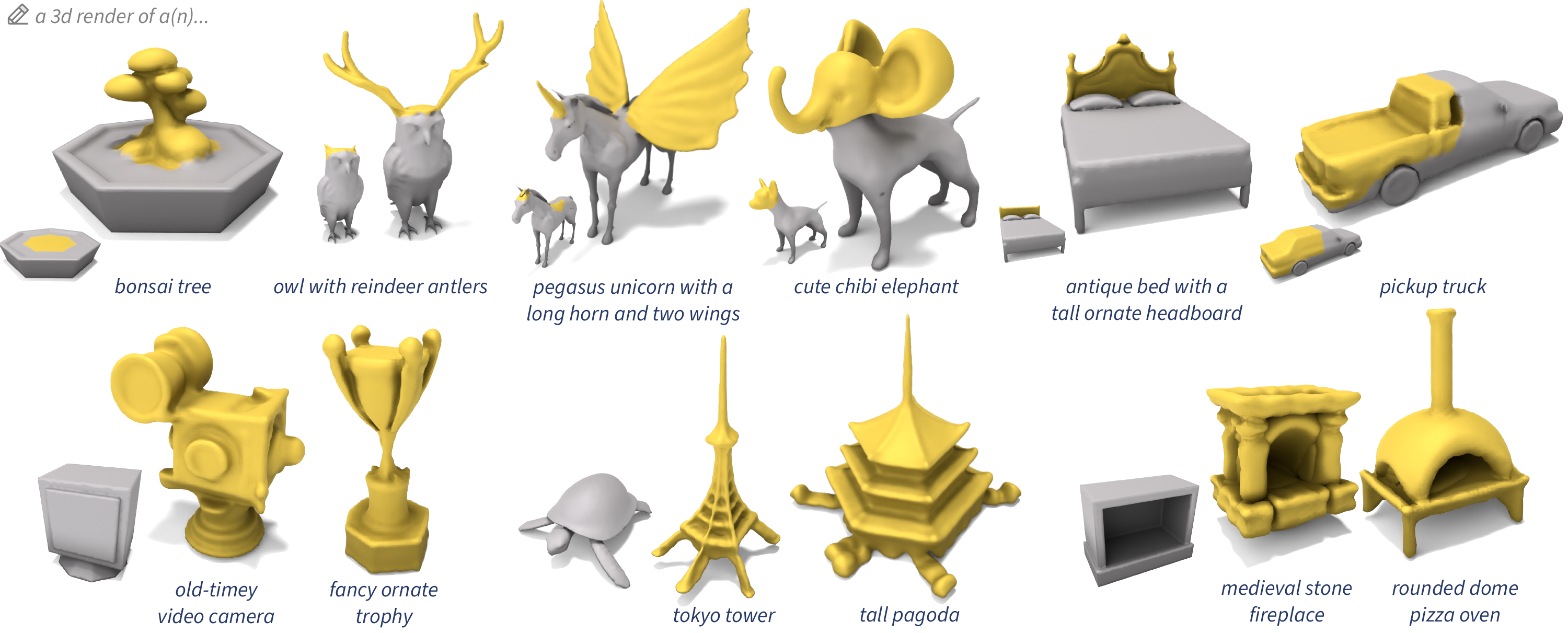}
    \caption{\textbf{Gallery of local growths and global detailing deformations}. \emph{Top row:} our method is capable of growing large parts out of small selected regions on diverse shapes, organic and manufactured. The results show symmetry, high detail even at extremities, and clean geometry at selection boundaries. \emph{Bottom row:} we can also perform global ``detailization'' deformations, where the whole shape is allowed to change. This is useful on simple base shapes and demonstrates the ability to change both high-level structure and finer details.}
    \label{fig:gallery}
\end{figure*}

Many recent methods optimize mesh deformations with visual losses  as guidance~\cite{gao2023textdeformer,kim2025meshup,dinh2025geomstyle}. These methods largely achieve their shape editing goals through finding vertex displacements and do not change mesh connectivity. While this approach is effective for a limited class of edits, there remains a wide class of shape deformations that are limited when keeping mesh connectivity constant. Growths, appendages, thin sheet-like volumes, and other large-scale structural reconfigurations generally cannot be achievable without excessive stretching or shrinking of triangles. Such edits may result in bad triangle quality, poor fidelity, and unsuitably small or large numbers of triangles forced to represent certain features of the shape.

We posit that \rh{such expressive deformations} in the fully explicit setting would strongly benefit from the ability to modify mesh connectivity, as connectivity alone implies some information about the underlying geometry. 
However, deformation editing methods have rarely incorporated remeshing, being inherently discrete in the required modifications of indices/pointers, making it challenging to involve discrete triangulation choices in the optimization (\eg \cite{rakotosaona2021diffdelaunay}).

Existing visual supervision-based remeshers \cite{palfinger2022continuousremeshing, barda2023roar} are meant for the inverse rendering task and are built with heuristics deeply tied to the use of \emph{vertex positions} as the optimization variable.
In contrast, the generative deformation task has been shown to greatly benefit from using differential deformation quantities as the optimization variable instead. These quantities, such as per-face Jacobians \cite{gao2023textdeformer, kim2025meshup} and per-vertex rotations \cite{dinh2025geomstyle}, solve into smooth, globally-informed deformations in the face of the noisy loss signals (namely SDS) used in the text-based, visually-supervised generative setting.

In our method, \emph{\ourmethod}, we propose a differential deformation optimization pipeline enhanced by remeshing, bringing the added expressiveness of connectivity modifications to the generative deformation task and its preferred deformation representations.

We extend the vertex normal-based deformation quantity of Geometry in Style \cite{dinh2025geomstyle}. As a vertex-based quantity, it allows straightforward transfer and interpolation from old mesh elements to new mesh elements during remeshing. Augmented with optimizable local scale in addition to local rotation, in tandem with remeshing at regular intervals, it enables growing large, detailed geometries. Deformation and remeshing can be localized to specific regions by construction, or globally applied across the mesh.
We employ an off-the-shelf optimizer (Adam \cite{kingma2014adam}) to update the deformation quantity, and interpolate its tracked state to the new discretization upon remesh. We adopt a standard isotropic remeshing algorithm, modified to enable this interpolation.

We demonstrate that this simple integration of remeshing together with our deformation quantity lets us achieve flexible shape optimizations that work well in generative settings with text-guided SDS (part growth, abstract shape detailing). Other methods that perform similar generative 3D tasks execute generation on dedicated proxy representations, \eg generating implicit shapes or multi-view image inpainting followed by 3D reconstructions. In contrast, we do not generate via alternative intermediate representations: we achieve desirable results by generative optimization \emph{directly on meshes}, using an image diffusion model. \rh{Our results highlight the effectiveness and versatility of a purely mesh-based generative optimization approach.}

\begin{figure}[t]
    \centering
    \includegraphics[width=0.99\linewidth]{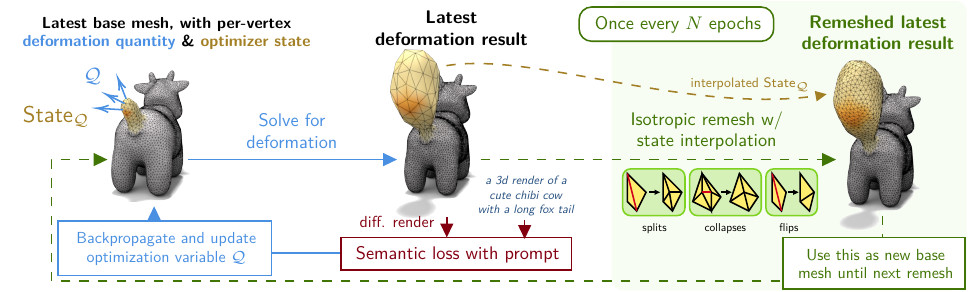}
    \caption{\textbf{Overview of our optimization loop.} Geometry is optimized via gradient descent of our deformation quantity $\QQ$, with associated optimizer state, with respect to a base mesh. At epoch $i$, we solve for the deformation using $\QQ_i$ (\cref{sec:deformqty}), render, take visual loss, and update the quantity. Once every $N$ epochs, we remesh the last deformed shape, interpolate the optimizer state onto the result, and use it as the new source mesh until the next scheduled remesh. (Note that the deformation in this figure is exaggerated for illustrative purposes; in practice, the remeshes are frequent enough that the deformation between each remesh is more subtle.)}
    \label{fig:overview}
\end{figure}

\section{Related Work} 
\label{sec:related_work}

\paragraph{Remeshing and triangle refinement.}
Classical geometry processing has a large body of work on remeshing and triangle refinement. We refer to \citet{khan2022remeshingsurvey,alliez2008recentadvances} for general surveys of remeshing methods for triangle meshes.
An overarching goal is uniformly-sized triangles, with no excessively small or large angles, numerically desirable for algorithms that are finite element discretizations, \eg those that use cotangent edge weights (which our method and others \cite{aigerman2022neural} fall under).

Remeshers belong to a few categories. One family of methods performs mesh parameterization followed by point sampling and Delaunay (or similar) triangulation on the parameter domain \cite{surazhsky2003isotropicparam, alliez2002interactiveremeshparam,alliez2003isotropicremeshparam,praun2003sphereparam,rakotosaona2021diffdelaunay}. Another family uses only local topology-preserving connectivity edits (edge collapses, splits, flips) to iteratively improve triangles across the mesh. Of these, particularly relevant are the works of Botsch \& Kobbelt \etal~\cite{botsch2004remesh} and Dunyach \etal~\citet{dunyach2013adaptive}, iterative methods for mesh improvement given a target edge length (either global or adaptive based on curvature). Edges are collapsed and split to maintain the target length range; edge flips and tangential smoothing improve vertex valence and triangle quality. This yields isotropic triangulations with near-equilateral triangle aspect ratios. Using these on-the-fly remeshing operations, editing \cite{dunyach2013adaptive, suzuki1998dragging} and correspondence \cite{maggioli2025rematching} paradigms have also been proposed. Similar operations have been built to simplify intrinsic triangulations \cite{liu2023surface} and non-manifold meshes \cite{liu2025simplifying}.

\fakepara{Vertex optimizers with remeshing.}
Recent applications of on-the-fly remeshing have included the inverse rendering and mesh reconstruction task, where remeshing is implemented directly in optimizers whose optimization variable is vertex positions. In particular, Palfinger~\cite{palfinger2022continuousremeshing} proposed a vertex optimizer and remesher that adapts Adam \cite{kingma2014adam} to take gradient statistics to determine and apply local remeshing operations (collapses, splits, flips). Connectivity is added in areas experiencing large change, making available more triangles for the geometry to evolve towards the inverse rendering target (\eg multiview images). Building on this work, Barda \etal~\cite{barda2023roar} improved heuristics and incorporated a raycaster; this has since been used for reconstructing meshes from other representations (\eg implicit fields \cite{barda2024magicclay} and multiview 2D inpainting \cite{barda2025instant3dit}). In contrast, our use of remeshing resembles that of \citet{nicolet2021largesteps,worchel2022reconeuraldeferred}, applying an isotropic remesher coarse-to-fine in a geometry optimization loop. Our vertex-based deformation quantity lets us remesh with interpolation of the geometry optimizer state, allowing more expressive and stable deformations than optimizing raw vertex displacements.

\fakepara{Generative mesh deformation.}
There is a sizable body of recent work in mesh deformation-based generative modeling of shapes. 
The seminal Neural Jacobian Fields method \cite{aigerman2022neural} has inspired a range of generative applications that use deformation for either primary generation or refinement of shapes. Methods with deformations as the main generative medium include TextDeformer \cite{gao2023textdeformer}, MeshUp \cite{kim2025meshup}, and Geometry in Style \cite{dinh2025geomstyle}. TextDeformer and MeshUp use Jacobians; Geometry in Style uses normals with differentiable As-Rigid-As-Possible (dARAP) solves,
though in a similar pipeline. In these methods, including our work, deformations (guided by a semantic visual loss) are the \emph{only} means of geometry production. Other work has used Jacobian deformation to create coarse geometry that is then textured \cite{yang2024dreammesh}, constrained creation of head avatars \cite{wang2025headdeformer}, improving a shape generated upstream \cite{Yu2025fancy123}, handle-based deformations \cite{yoo2024apap,baieri2024implicitarap}, and shape correspondence \cite{sundararaman2024deform,liu2024multiax}.

These studies are part of the wider literature on shape generation, which has seen a proliferation of methods, particularly implicits, images, or latent representations followed by mesh reconstruction or improvement \cite{lai2025hunyuan3d,li2024craftsman3d,Xiang2025structuredlatents,xu2024instantmesh,sun2024recentimplicits,zhang2023vecset,yan2025omages,zhang2024clay,siddiqui2024meta,wang2023prolificdreamer,jincheng2025craftmesh,chen2025artdeco,liang2024luciddreamer,zhu2023hifa,jung2023meshdensityadapt,Erkoc2025preditor3d,palandra2024gsedit}, and autoregressive sequence models of discrete elements \cite{Zhao2025deepmeshArtistmesh, tang2024edgerunner,Wang2025nautilus,hao2024meshtron,Siddiqui2024meshgpt,Li2025meshpad,gao2025mars,zhang2025geomdistr}. Many of these methods have been enabled by advances in lifting 2D visual priors to 3D in a variety of generative and analysis tasks and representations \cite{dreamfusion,scorejacobian,clip,michel2022text2mesh,fantasia3d,magic3d,scorejacobian,seo2023let,efficientdreamer,sweetdreamer,gaussiansplatting,zero123,zhang2023text2nerf,clipnerf,decatur2023highlighter,decatur2024paintbrushcsd}.

\begin{figure}[t]
    \def\wdth{0.99}
    \centering
    \includegraphics[width=\wdth\linewidth, trim=0 25 0 0 clip]{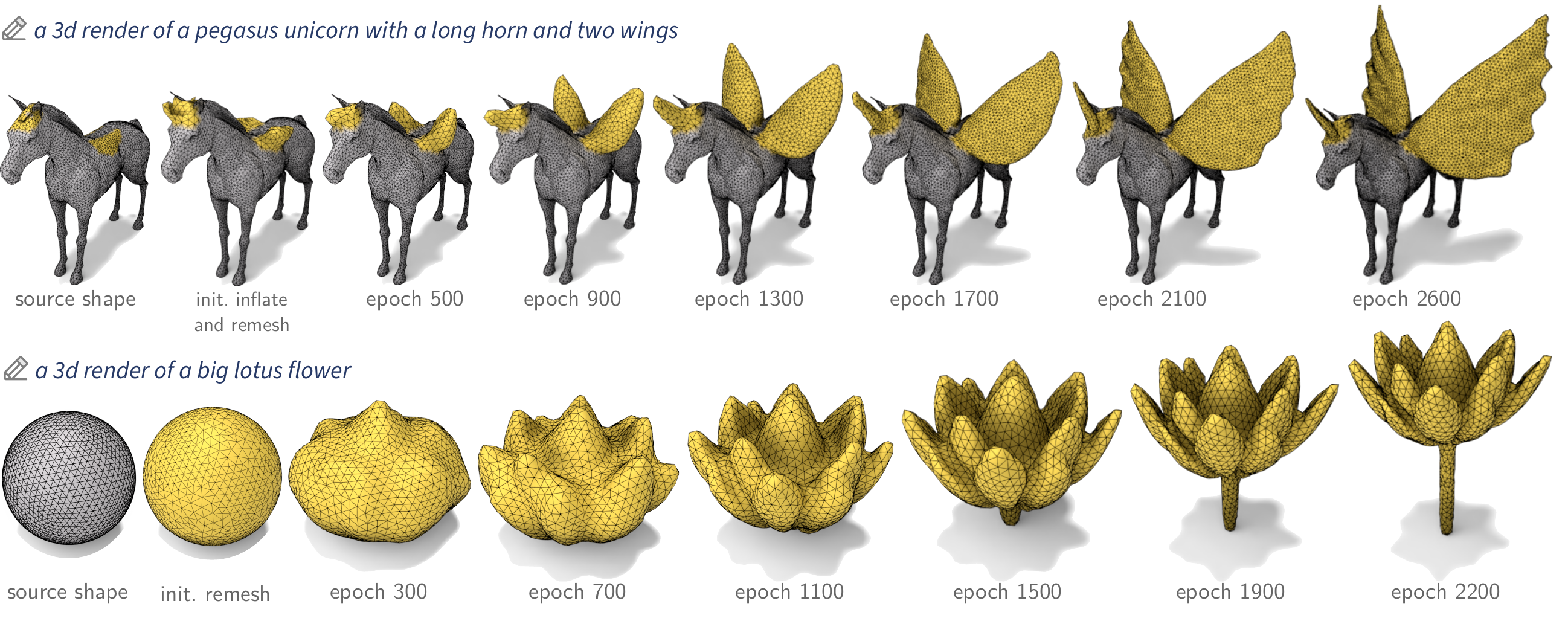}
    \caption{\textbf{Evolution.} We visualize the geometric evolution through the course of two example optimizations, one localized and the other a global deformation. The local growth example involves disjoint selected components (the wings, the horn), which all grow simultaneously. The global deformation shows an ability to form complex geometric detail (starting from a shape as basic as a sphere) and appendages while maintaining clean isotropic triangulation throughout.}
    \label{fig:evolution}
\end{figure}

\section{Method} \label{sec:method}

\subsection{Overview}
Given an original source mesh $\mathcal M_0$ with vertices and faces $(\mathcal V_0, \mathcal F_0)$, we wish to \textit{deform and remesh} $\mathcal M_0$ to find a new mesh $\mathcal M^*=(\mathcal V^*, \mathcal F^*)$ that 
meets a target text prompt $\mathbf x$.
A selection region of vertices can optionally be specified, represented as a binary selection mask ($\mathsf{sel}(k)=1$ denotes vertex $k$ is part of the selection, and $\mathsf{sel}(k)=0$ denotes $k$ is not selected), where deformation and remeshing are meant to be restricted to selected vertices only.

\cref{fig:overview} presents an overview of the optimization loop and our use of remeshing. The desired mesh $\mathcal M^*$ is constructed via gradient descent optimization; at any given iteration or epoch $i$, this loop's optimization variable is a \textit{deformation quantity} $\caltil Q_i$ defined with respect to the loop's \textit{current base mesh} $ \caltil M^{r_i} = (\caltil V^{r_i}, \caltil F^{r_i})$ where $r_i$ denotes the current number of remeshes done thus far at epoch $i$. 
\begin{wrapfigure}{l}{0.55\linewidth}
    \centering
    \includegraphics[width=1.02\linewidth,trim=0 0 0 0 clip]{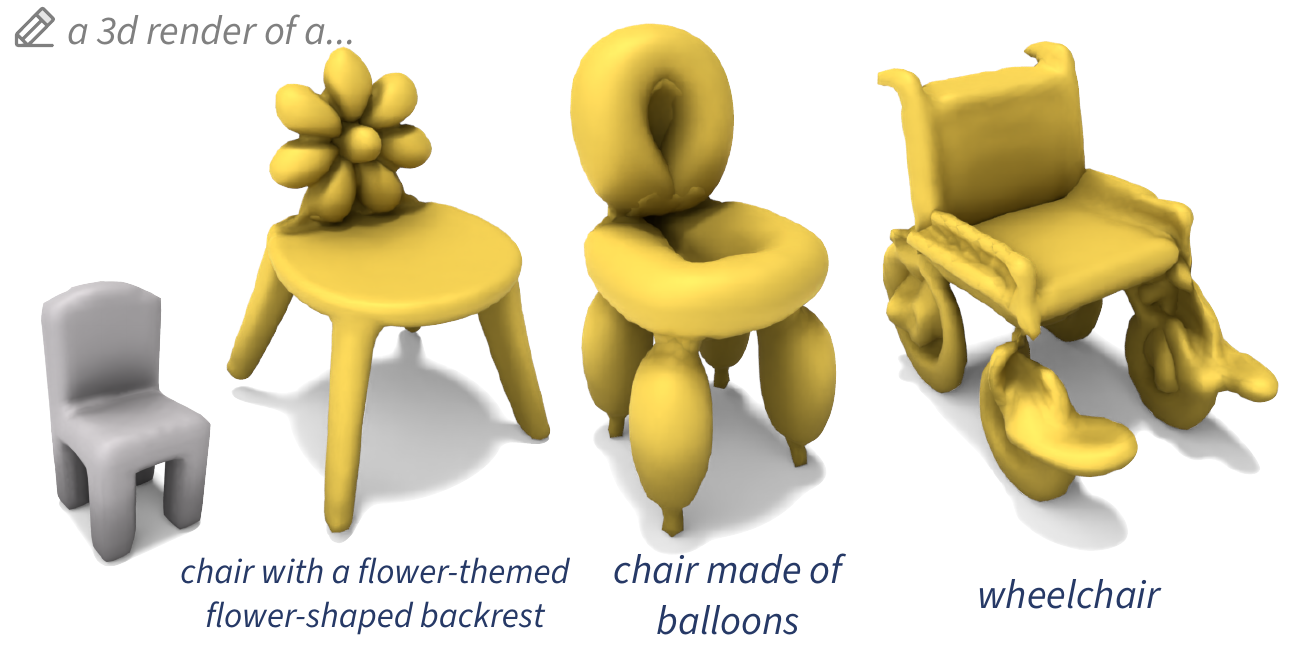}
    \caption{\textbf{Global detailization}. Our method is able to take basic shapes and globally deform \& remesh them expressively towards prompts. The results keep high-level structure (\eg four similar legs) while exhibiting expressive geometric detail: the folds on the \textit{flower-themed flower-shaped} backrest, the tapered lobes of the \textit{chair made of balloons}, the \textit{wheelchair}'s wheels.}
    \vspace{-5pt}
    \label{fig:chairs}
\end{wrapfigure}
The \textit{current base mesh} is initialized to $\mathcal M^0$, which is the original source mesh $\mathcal M_0$ with preprocessing applied (described in \cref{sec:preproc}). 

At optimization epoch $i$, a deformation function $D$ finds deformed vertices $\caltil V'_i$ using the latest value of the deformation quantity $\caltil Q_i$ together with the associated base mesh $\caltil M^{r_i}$ for which it is defined, \ie $\caltil V'_i := D(\caltil M^{r_i}, \caltil Q_i)$. This deformation gives the shape to be rendered for visual losses, and keeps the current base triangulation $\caltil F^{r_i}$ fixed. The deformation formulation is described in \cref{sec:deformqty}.

The deformed mesh $(\caltil V'_i, \caltil F^{r_i})$ is then rendered with a differentiable renderer, and the images fed to a cascaded score distillation (CSD) loss providing the semantic supervision to be backpropagated to update the deformation quantity. This gives $\caltil Q_{i+1}$ for the next epoch (epoch $i+1$); see \cref{sec:csd}.

At regular intervals in this optimization loop, we interrupt the geometry optimization and perform a remesh using the remeshing function $R$. The remesh will \textit{replace} the current base mesh with a remesh of the latest deformation result. More precisely, let $\mathsf{do\_remesh}(i)=1$ mean that a remesh is scheduled for optimization epoch $i$. For each $i$ such that $\mathsf{do\_remesh}(i)=1$, we compute the \emph{next base mesh} $\caltil M^{r_i+1} := R(\caltil V'_i) = R(D(\caltil M^{r_i}, \caltil Q_i))$. We describe the remeshing method $R$ in \cref{sec:remesh}. 
This leads to $\caltil Q_{i+1}$ becoming reinitialized, since the base mesh $\caltil M^{r_i}$ that $\caltil Q_i$ was associated with has now been switched out for $\caltil M^{r_i+1}$. 
Crucially, we also interpolate and carry forward optimization states in order to then resume optimizing geometry after the remesh; we describe this in \cref{sec:drmsh}.

\begin{figure}[t]
    \begin{subfigure}[t]{0.5\linewidth}
        \includegraphics[width=0.99\linewidth,trim=0 0 0 0 clip]{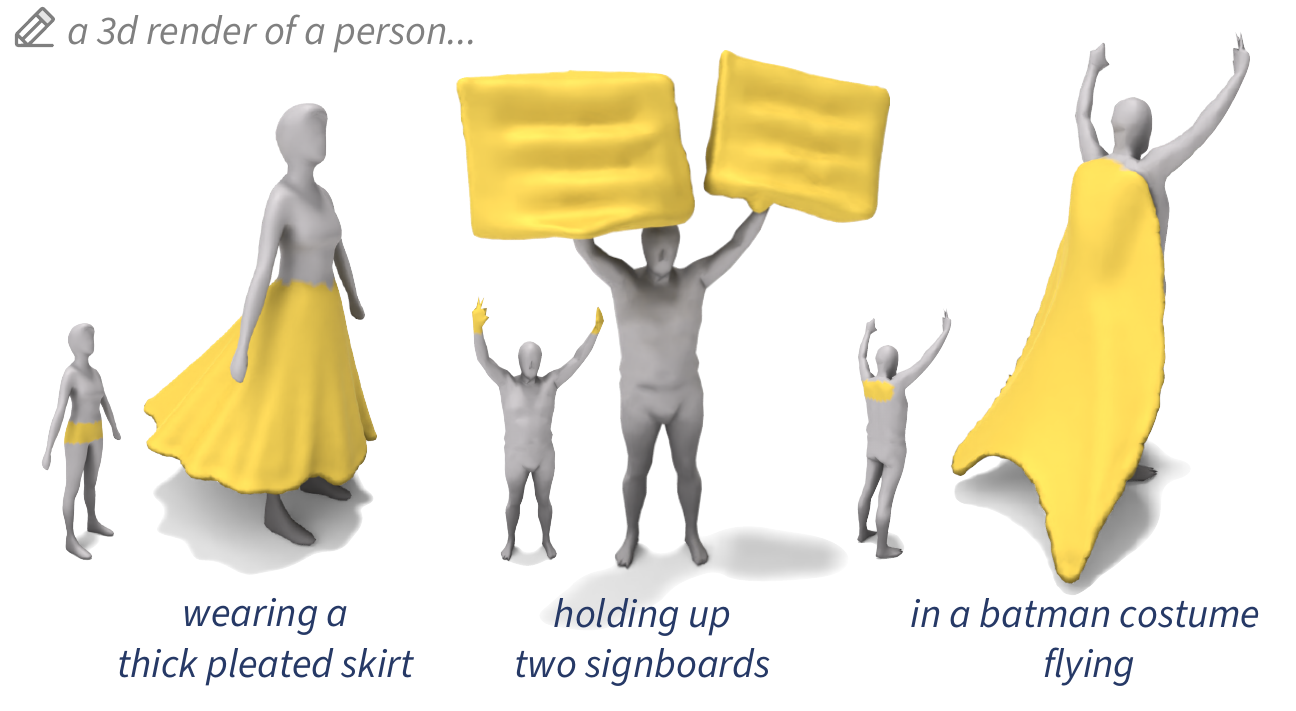}
        \caption{\textbf{Growing garments}}
        \label{fig:humans-garments}
    \end{subfigure}
    \begin{subfigure}[t]{0.5\linewidth}
        \includegraphics[width=0.93\linewidth]{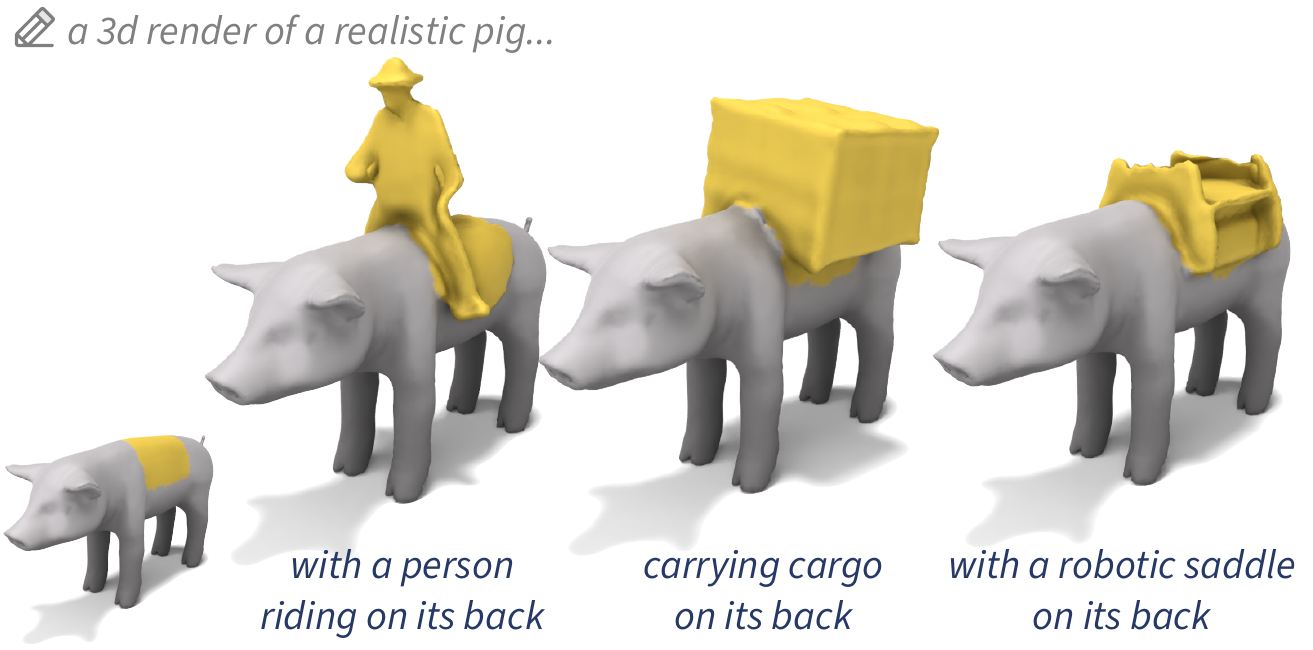}
        \caption{\textbf{Same selection, different growths}}
        \label{fig:pigs}
    \end{subfigure}
    \caption{ \textbf{More local growth examples.} \textbf{(a) Growing garments:}  Our method is able to generate thin parts and garments from small selection regions on human shapes, extending considerably outwards from the original extents with high detail at the generated extremities. \textbf{(b) Same selection, different growths:}  On the same selection and mesh, our method can deform \& remesh to produce varied growths, such as the \emph{person} riding (including a detailed hat), the \emph{cargo} with geometric texture on the surface, and the \emph{robotic saddle} with sharp features.}
\end{figure}

\subsection{Deformation Quantity and Method}
\label{sec:deformqty}
The deformation quantity $\mathcal Q$ we use is a 6-component vector for each vertex, \ie $\mathsf q_k \in \mathbb R^6$ for each vertex $k$, and so $\QQ= \{\qq_k \in \mathbb R^6 \mid k \in \{1 \dots |\VV|\}\}$. The first three features, which we denote by $\qq_k^\mathrm{dir} \in \mathbb R^3$, describe a direction vector, and the last three features, denoted by $\qq_k^\mathrm{scale} \in \mathbb R^3$, describe a scaling vector for the three world axes. 

This deformation quantity is an extension of the per-vertex $\mathbb R^3$ normal vector quantity of Geometry in Style \cite{dinh2025geomstyle}. Originally proposed for its inherent regularization using differentiable ARAP, its vertex-based deformation quantity is also advantageous in a remeshing setting. Our remesh operations operate with vertex attributes; thus, a vertex-based optimization quantity makes it straightforward to interpolate optimizer state onto new mesh elements during a remesh. 
This contrasts with the more conventional per-face Jacobians \cite{aigerman2022neural}, a face-based quantity that would be more lossy to cast onto vertices and interpolate during a remesh.
In order to support expressive deformations and growing large parts beyond the extents of the original selection (see ablation in \cref{fig:noscaleabl}), we append features to Geometry in Style's representation that encode scale in addition to rotation. Accordingly, the deformation function $D$ is
\begin{equation}
D(\MM, \QQ) = \mathrm{GlobalStep}(\MM, \mathrm{LocalStep}(\MM, \QQ))\\
\end{equation}
The local step first finds a local transform $\Tt_k$ for each vertex $k$,  composed of a rotation matrix $\Rr_k$ computed using $\qq_{k}^\mathrm{dir}$ and a diagonal scaling matrix $\Ss_k$ formed from $\qq_k^\mathrm{scale}$. The global step then solves for global vertex positions that best satisfy those local transformations.

The rotation is found by \cref{eq:procrustes}, a Procrustes problem, which finds a rotation matrix $\Rr_k$ taking the set of vectors $\NN_k \cup \{\uu_k\}$ to the set of vectors $ \NN_k \cup \{\qq_k^\mathrm{dir}\}$, where $\uu_k$ is the normal of vertex $k$ in the current base mesh and $\NN_k$ is the set of edge vectors in the spokes-and-rims neighborhood of vertex $k$. Here, $\pp_i,\pp_j$ are the position vectors of vertices $i$ and $j$; $w_{ij}$ is the cotangent weight of edge $(i,j)$; $a_k$ is the Voronoi area of vertex $k$; and the hyperparameter $\lambda$ controls the strength of the rotation towards the goal $\qq_k^\mathrm{dir}$, as described in \citet{dinh2025geomstyle} and \citet{liu2021normal}.
{\small
\begin{align}
\begin{split}\label{eq:procrustes}
&\Rr_k = \argmin_{\Tilde{\Rr}_k \in \SO(3)}
\sum_{(i,j) \in \NN_k} w_{ij} \| \Tilde{\Rr}_k (\pp_j - \pp_i) - (\pp_j - \pp_i) \|_2^2
\\
&\hspace{1.8cm}
+ \lambda a_k \| \Tilde{\Rr}_k \uu_k -  \qq_{k}^\mathrm{dir} \|_2^2
\end{split}
\end{align}
}\vspace{-1.3em}
\begin{align}
\Ss_k &= \mathrm{diag}\left(\qq_k^\mathrm{scale}\right) \label{eq:scalemat}\\
\Tt_k &= \Ss_k \Rr_k\\
\mathrm{LocalStep}(\MM, \QQ) &= \{ \Tt_k \mid k \in \{1\dots|\VV|\} \}
\end{align}
The global step is the differentiable ARAP global solve from Geometry in Style except for the use of our local step's composed transformations $\TT = \{\Tt_k \mid k \in \{1\dots|\VV|\}$ rather than just the rotations $\{\Rr_k \mid k \in \{1 \dots |\VV|\}\}$.
{\small
\begin{align}
\begin{split}\label{eq:global}
&\mathrm{GlobalStep}(\MM,\TT) \\ &=\argmin_{\VV'}\sum_{k \in \{1\dots |\VV|\}} \sum_{(i,j) \in \NN_k} w_{ij}\| \Tt_k (\pp_j - \pp_i) -  (\pp_j'-\pp_i') \|_2^2
\end{split}
\end{align}
}
\paragraph{Restricting deformation to a selected region.}
\label{sec:selectiondeform}
To restrict the deformation to the selected region, we 1) set $\Tt_k$ to be identity for $k$ where $\mathsf{sel}(k)=0$ and 2) substitute the original positions of these vertices into \cref{eq:global} before solving for the unknowns. The solutions are thus the deformed positions for only the selected vertices. We describe the matrix form of this solve in the supplementary material.

\fakepara{Designing constraints on deformations.}~ Having separate $\qq_k^\mathrm{scale}$ and $\qq_k^\mathrm{dir}$ components (scale \& direction) on the deformation quantity lets us constrain the possible deformations \emph{by construction} (rather than by losses). For global deformations, we apply a soft clamp (leaky ReLU) on the $\qq_k^\mathrm{scale}$ input to \cref{eq:scalemat} with a leaky ReLU floor of $0.98$. This combats the supervision model's tendency towards shrinkage when the whole mesh is optimizable; see supplementary material for ablation.

\fakepara{Initialization.} Upon a new base mesh, for vertex $i$, $\qq_i^\mathrm{dir}$ is initialized to the area-weighted vertex normal, and $\qq_i^\mathrm{scale}$ is initialized to all 1.

\subsection{Remeshing}
\label{sec:remesh}

We use the remeshing method of Botsch \& Kobbelt \citet{botsch2004remesh}. As an isotropic remesher that aims to produce uniform, near-equilateral triangles given a target edge length, one iteration of the method involves 1) an edge collapse step, 2) an edge split step, 3) a valence-improving edge flip step, 4) a tangential smoothing step, and 5) a projection step to project the result of these steps to the original surface. These steps make up one iteration; many such iterations can be done for better isotropic triangulations. In practice, we perform 2 such iterations once every 100 deformation epochs.

In propagating the vertex selection, we incorporate the selection flag $\mathsf{sel}(k)$ as a real-valued vertex attribute $\mathsf{sel}_\mathbb R(k)$. All remeshing operations (edge collapse, edge flip, edge split) only occur on edges $(u,v)$ where $\mathsf{sel}_\mathbb R(k) \geq 0.5$ for both $k \in \{u,v\}$. Edge collapses and splits grant the resulting vertex $w$ with $\mathsf{sel}_\mathbb R(w) = \frac12\left(\mathsf{sel}_\mathbb R(u) + \mathsf{sel}_\mathbb R (v)\right)$. After a complete remesh iteration (\ie after reprojection), every $\mathsf{sel}_\mathbb R(k)$ is rounded to $1$ if at least $0.5$, and $0$ otherwise.

The Botsch \& Kobbelt method requires a target length for each edge. We use a global target length for all edges involved in remeshing. 
This length is on a coarse-to-fine schedule, critical for substantial growth of geometry as shown
in \cref{fig:remeshonceablation}.
We describe our schedule for the local and global tasks in \cref{sec:experiments}.

Despite the connectivity change, we retain correspondences with the original mesh. We keep perfect face and vertex correspondence outside modified regions (see \cref{fig:uvtex}). Within modified regions, vertex attributes are interpolated through; the optimizer state is one such attribute (\cref{sec:drmsh}).

\begin{figure}[t]
    \centering
    \includegraphics[width=0.99\linewidth, trim=0 20 0 0 clip]{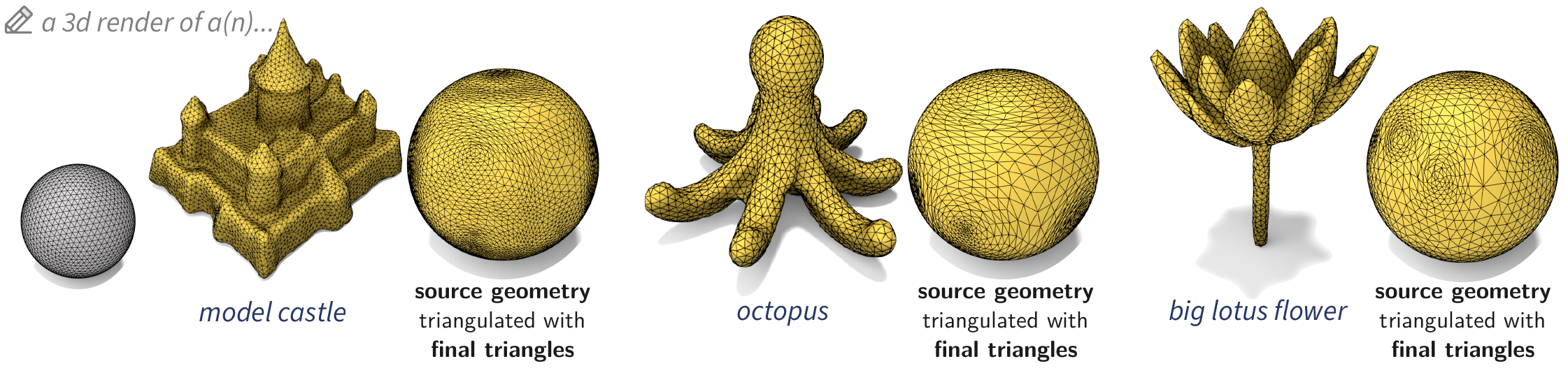}
    \caption{\textbf{Generated triangulation tailored for the generated geometry.} Taking global deformation of a sphere as an example, we show that each final triangulation, applied to the original geometry, encodes the structures and appendages requiring high vertex density in the result shape. This shows the appropriateness and adaptivity of the generated triangulation for the generated geometry, not merely a uniform upsample of the original triangulation.}
    \label{fig:spheres}
\end{figure}

\subsection{State Interpolation for Resuming Optimization}
\label{sec:drmsh}
We note that the reprojection step involves computing barycentric coordinates on the pre-remesh triangles for placing new vertices. We take advantage of this to interpolate the \emph{optimizer state}. In most optimizers, including Adam\cite{kingma2014adam} which we use, the state is an accumulator of the gradient with respect to the variable. Since our deformation quantity is vertex-based, the optimizer state is also vertex-based and can be easily interpolated with barycentric coordinates.
In the supplementary and \cref{fig:interpabl}, we show that this interpolation is important for stable, symmetric growths. In global deformations, it also prevents unchecked scale-ups leading to excess resolution without better geometry.

\subsection{Preprocessing}
\label{sec:preproc}
We normalize all starting meshes such that their axis-aligned bounding box fits within the standardized cube $[-1, 1]^3$ in world coordinates.
After normalization, for the local growth task in particular, we apply \emph{initial inflation} to provide a starting geometry, which SDS supervision is sensitive to. 
To each initial selected vertex we add a displacement of length $\ell$ along its vertex normal direction. We find $\ell$ by a heuristic formula based on the sum volume of the selected patches, computed after closing holes making them watertight surfaces:
\(
\ell = a^{-b \sqrt[3]{V}}
\)
where $V$ is the sum volume of all hole-closed selection patches. For positive $a, b$, this formula grants larger inflation to smaller selections, the relation decaying exponentially. (See supplementary material for an ablation on this initial inflation step.) Preprocessing concludes with a first remesh of the base geometry before the optimization loop begins.

\subsection{Differentiable Rendering \& Supervision}
\label{sec:csd}

To obtain supervision for the deformation quantity, we follow the common protocol of differentiable rendering into images followed by a 2D image diffusion model providing score distillation sampling (SDS) loss. We use Cascaded Score Distillation \cite{decatur2024paintbrushcsd}, an SDS variant making use of the multiple diffusion stages of DeepFloyd IF \cite{stabilityai2023deepfloydif}. We use \texttt{nvdiffrast} \cite{Laine2020diffrast} as our differentiable renderer.

\section{Experiments} \label{sec:experiments}

\paragraph{Hyperparameters.}
Each remesh consists of 2 Botsch \& Kobbelt iterations. A remesh happens every $N=100$ epochs. Local growth runs last at most 2600 epochs, and global deformations 2200 epochs. The initial inflation (\cref{sec:preproc}), used only for local growths, has $a,b$ empirically set to $a=4.3$, $b=5.2$. We use $\lambda=8$ for \cref{eq:procrustes}. Each epoch uses 8 views sampled uniform-randomly from fixed ranges for azimuth, elevation, and FOV (settings are shared across all runs within each task: local growths, global deforms). See supplementary material.

\def\degree{^\circ}
\fakepara{Target length schedule.}~~For local growths, the coarse-to-fine target length schedule begins at $1.7\times$ the average edge length of the mesh after all preprocessing, linearly ramping from $1.7\times$ to $1.3\times$ over the next 10 remeshes, then from $1.3\times$ to $1.0\times$ over the next 5 remeshes, then $1.0\times$ for the remainder of remeshes. Global deformations begin at $1.4\times$ the average edge length of the preprocessed mesh, ramp to $1.0\times$ over the next 18 remeshes, and set to $1.0\times$ for the remainder. We show the significance of this coarse-to-fine schedule as opposed to remeshing finely once or remeshing curvature-adaptively in \cref{fig:remeshonceablation,fig:adaptivermshabl}.

\subsection{Qualitative Evaluation}
\paragraph{Local growths: large extents \& clean detail.}
We demonstrate our method's ability to impart large-scale yet detailed growth of whole parts to even small selections on a variety of shapes, organic and manufactured. The growths extend far beyond the source selection's spatial extent, such as the \emph{long lizard tail}, the \emph{two pegasus wings} (\cref{fig:teaser}), the large \emph{cargo} on the pig's back (\cref{fig:pigs}), the large \emph{reindeer antlers} (\cref{fig:gallery}), and the skirt and cape growths (\cref{fig:humans-garments}).

They also exhibit expressive geometry even at the far ends of growths: the antlers' branching tines (\cref{fig:gallery,fig:comparisonbig}), the curled trunk of the \emph{elephant} (\cref{fig:gallery}), the feather texture on the \emph{wings} examples in \cref{fig:gallery,fig:teaser}, the ``text'' on the signboards in \cref{fig:humans-garments}, and a hat on the growth of a person riding on a pig's back in \cref{fig:pigs}. Noteworthy is also the ability to ``remove'' and reconfigure structures: in the \emph{bowler hat} example (\cref{fig:teaser}) the horns dissolve into a hat; in \cref{fig:gallery} the back of the car shrinks to become the \emph{pickup truck}'s bed.

\fakepara{Expressive global detailization.}~ When the selection is the entire shape, our method generates highly detailed geometry out of plain, simple shapes, closely matching the text prompt. \cref{fig:chairs} demonstrates the deformation of a chair subject to different prompts. While the source chair has simple geometry, it is successfully deformed and remeshed into significantly detailed, distinct outputs. For example, the backrest turns into a crisp flower geometry, handles and wheels emerge, or balloon-like elements form. Even with drastic changes, the chair's semantic structure (backrest, seat, four legs) is preserved.

\begin{center}{
    \def\wdth{0.99}
    
    \centering
    \includegraphics[width=\wdth\linewidth]{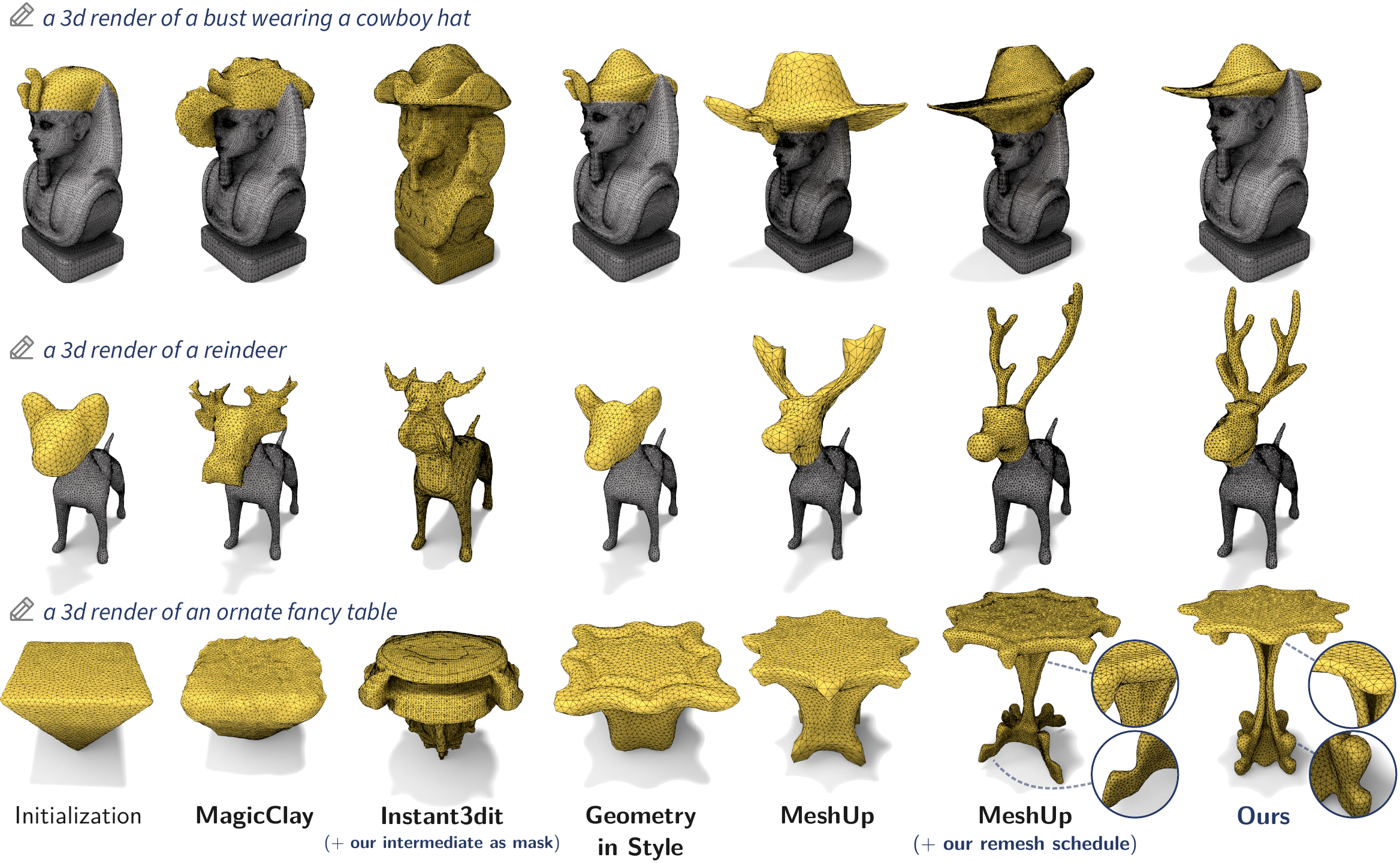}
    \vspace{-1mm}
    \captionof{figure}{\textbf{Qualitative comparison}. We compare our method with recent mesh deformation techniques \cite{barda2024magicclay,barda2025instant3dit,dinh2025geomstyle,kim2025meshup} for both local growths and global detailization. The baselines produce noisy results (MagicClay and Instant3dit) or are limited in the generated geometry expressivity (Geometry in Style and MeshUp). Our remeshing schedule also helps MeshUp, though our method still produces cleaner, symmetric deformations, with more efficient use of resolution (see insets), aligning better with the source mesh. Note that Instant3dit requires a 3D mask to inpaint; we use an early epoch from our run as this mask, plus 0.25 dilation, since  the selection mask alone restricts growth-like inpainting.
    Instant3dit's public code does not include the adaptive remesher or closed-source mesh LRMs from the paper.}
    \vspace{2mm}
    \label{fig:comparisonbig}
}\end{center}Taking global detailization further, in \cref{fig:spheres} we showcase the deformation of a primitive source shape as simple as a sphere. Beyond the expressive outputs, this experiment stresses the specificity of our remeshing process: rather than a mere subdivide or upsample of the source geometry, triangles appear \emph{as needed} to form the desired outcome.
This corresponds to an efficient, \textit{non-uniform} triangulation when viewed on the original geometry: triangle density corresponds to the \emph{model castle}'s towers, the \emph{octopus} tentacles, and the \emph{lotus flower}'s receptacle and petals, and is sparser elsewhere. 
We attribute this to the frequent coarse-to-fine remesh schedule, accompanying the geometry optimization as it forms new parts and details. Our optimizer state interpolation also helps: the \citet{botsch2004remesh} remesher aims to get all edges to a target length, and when growth momentum is carried through remeshing via our optimizer state interpolation, actively growing regions (edges becoming longer) also get more triangles as needed.

{\def\putcaptionhere{\caption{\textbf{Quantitative comparison}: CLIP and VQA scores ($\times 100$, higher is better), average face quality (FQ) ($\times 100$, higher is better), average face counts (\#F). We evaluate on $n=64$ shape-prompt pairs (see supplementary material for details).
    Our method achieves superior CLIP and VQA score, suggesting better perceptual quality and adherence to prompts, as well as superior triangle quality while being efficient with face budget among the methods that remesh. (*Geometry in Style and MeshUp do not remesh; their average face counts thus reflect the resolutions of the source meshes.) 
    }}
    \setlength{\tabcolsep}{2pt}
    \begin{table}[t]
        \putcaptionhere
        \small 
        \renewcommand{\arraystretch}{1.1}
        \centering
        \begin{tabular}{l@{~~~~}c@{~~~~}c@{~~~~}c@{~~~}c@{~~}}
        \toprule
            Method & CLIP$\uparrow$ & VQA$\uparrow$ & FQ$\uparrow$ & \#F \\
        \midrule
            MagicClay \cite{barda2024magicclay} & 29.821 & 57.150 & 86.784 & 17203.55 \\
            Instant3dit \cite{barda2025instant3dit} & 28.399 & 52.111 & 80.132 & 40034.69 \\
            Geometry in Style \cite{dinh2025geomstyle}& 29.279 & 52.120 & 94.146 & 11308.06* \\
            MeshUp \cite{kim2025meshup}& 30.725 & 66.093 & 84.831 & 11308.06* \\
            MeshUp \textbf{+ our~remesh}& 31.148 & 69.360 & 96.310 & 29461.34 \\
            \textbf{\ourmethod\ (\textbf{ours})}& \textbf{31.416} & \textbf{72.930} & \textbf{96.516} & {16963.75} \\
        \bottomrule    
        \end{tabular}
        \label{tab:numbers}
    \end{table}
}

\subsection{Comparison}

We contrast \ourmethod{} with some recent generative deformation baselines \cite{barda2024magicclay,barda2025instant3dit,dinh2025geomstyle,kim2025meshup}. \cref{fig:comparisonbig} presents visuals; \cref{tab:numbers} reports numeric results on 64 shape-prompt pairs. Rather than defining the deformation objective on the mesh surface, MagicClay and Instant3dit generate on intermediate proxy representations: implicit SDFs and multiview image inpainting, respectively. These representations are lossy, more detached from the formation of a mesh, resulting in artifacts. In contrast, Geometry in Style and MeshUp deform the mesh directly. However, both lack remeshing and are not expressive enough. Our method enjoys both worlds: it deforms and remeshes the mesh itself, generating highly detailed and clean geometries. Notably, adding our remesh schedule to MeshUp boosts expressivity, showing the benefit of remeshing in explicit mesh deformation. 

Quantitatively we evaluate via four metrics: CLIP score \cite{clip} measuring CLIP similarity of result renders to the text prompt; VQA score \cite{lin2024vqa} as a perceptual score for text-to-visual generative models; average triangle face quality (FQ) defined as $4\sqrt3\frac{A}{\ell^2_1 +\ell^2_2+\ell^2_3}$ \cite{shewchuk2002good,bhatia1990fem} (face area $A$, edge $i$ length $\ell_i$) where equilateral triangles achieve the ideal score of 1; and face count, measuring triangulation efficiency. Among compared methods, ours achieves the best CLIP and VQA scores, indicating perceptual quality and fidelity to the text prompt. Since we regularly run isotropic remeshing, we maintain higher quality triangles than competing methods, which either only distort triangles without remeshing, or reconstruct towards proxy representations, not ensuring good elements. Our quality is also more efficiently achieved, with a lower face count than the baselines that remesh. 

\subsection{Ablation Study}

\begin{figure}
    \centering
    \begin{subfigure}[b]{0.255\linewidth}
        \centering
        \includegraphics[width=0.99\linewidth,trim=0 -1.8em 0 0]{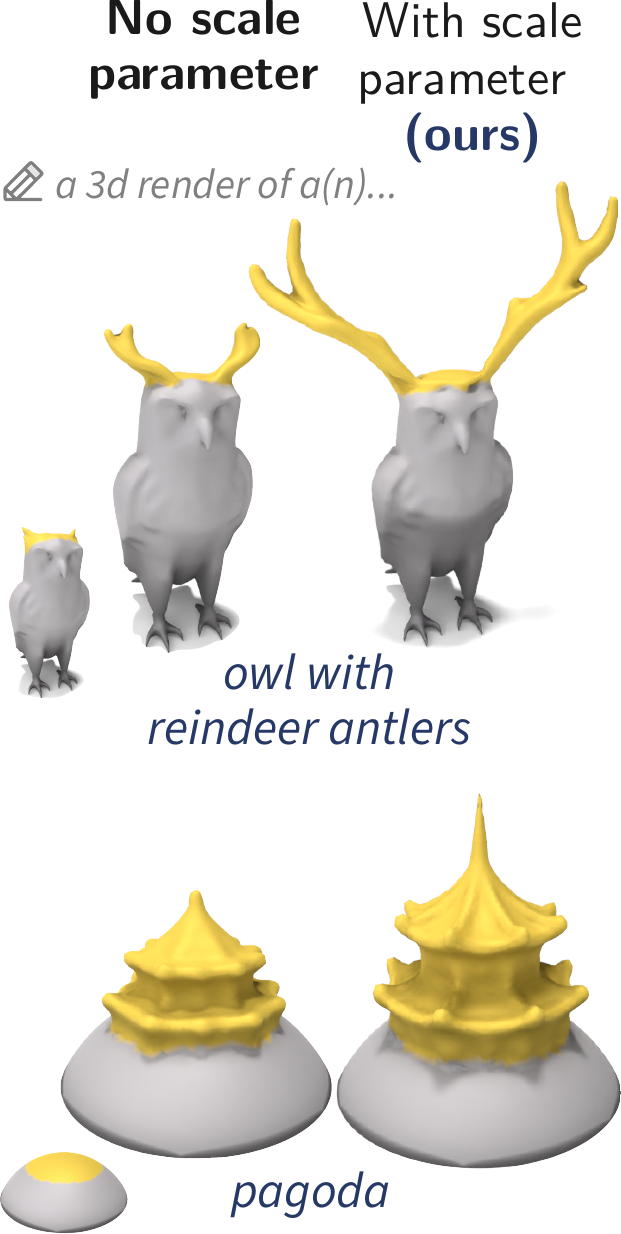}
        \caption{\centering\textbf{Ablation:\\no scale parameter}}
        \label{fig:noscaleabl}
    \end{subfigure}
    \qquad
    \begin{subfigure}[b]{0.62\linewidth}
        \includegraphics[width=0.99\linewidth,trim=0 -2mm 0 0]{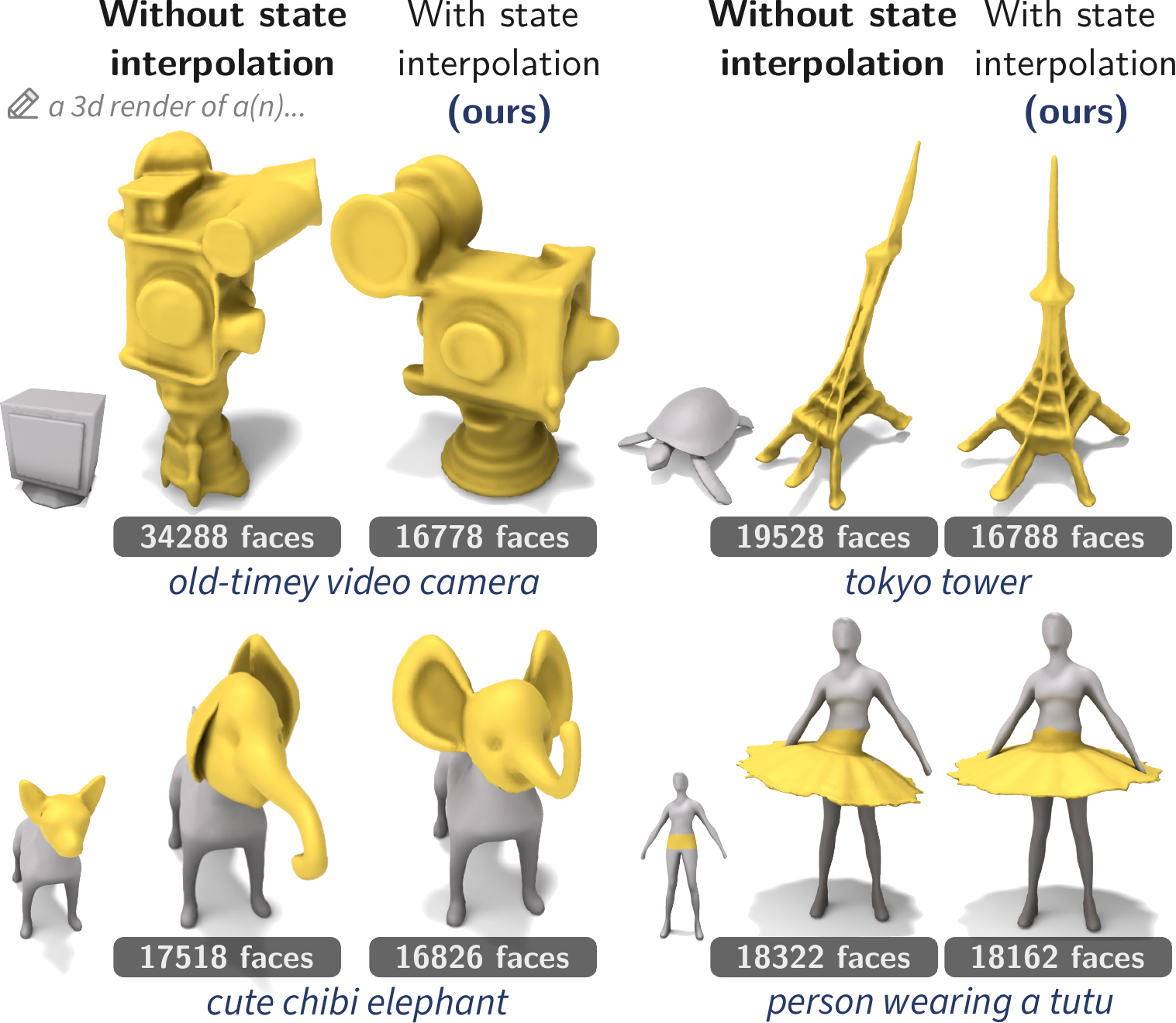}
        \caption{\centering\textbf{Ablation: no optimizer state interpolation.}}
        \label{fig:interpabl}
    \end{subfigure}
    \caption{\textbf{Ablations.}
    \textbf{(a) No scale parameter:}  Without the optimizable scale parameter, growths fail to attain large extents and have blunted, less sharp detail at the extremities.
    \textbf{(b) No optimizer state interpolation:}  Without optimizer state interpolation, the optimization loses all gradient history upon each remesh. This leads to less stable deformations, asymmetries, exaggerated view-dependent artifacts, and, in some cases, excessive resolution or scale-up without better geometry.
    }
\end{figure}

\begin{figure}[t]
    \centering
    \begin{subfigure}[b]{0.69\linewidth}
        \includegraphics[width=0.99\linewidth, trim=0 0 0 0 clip]{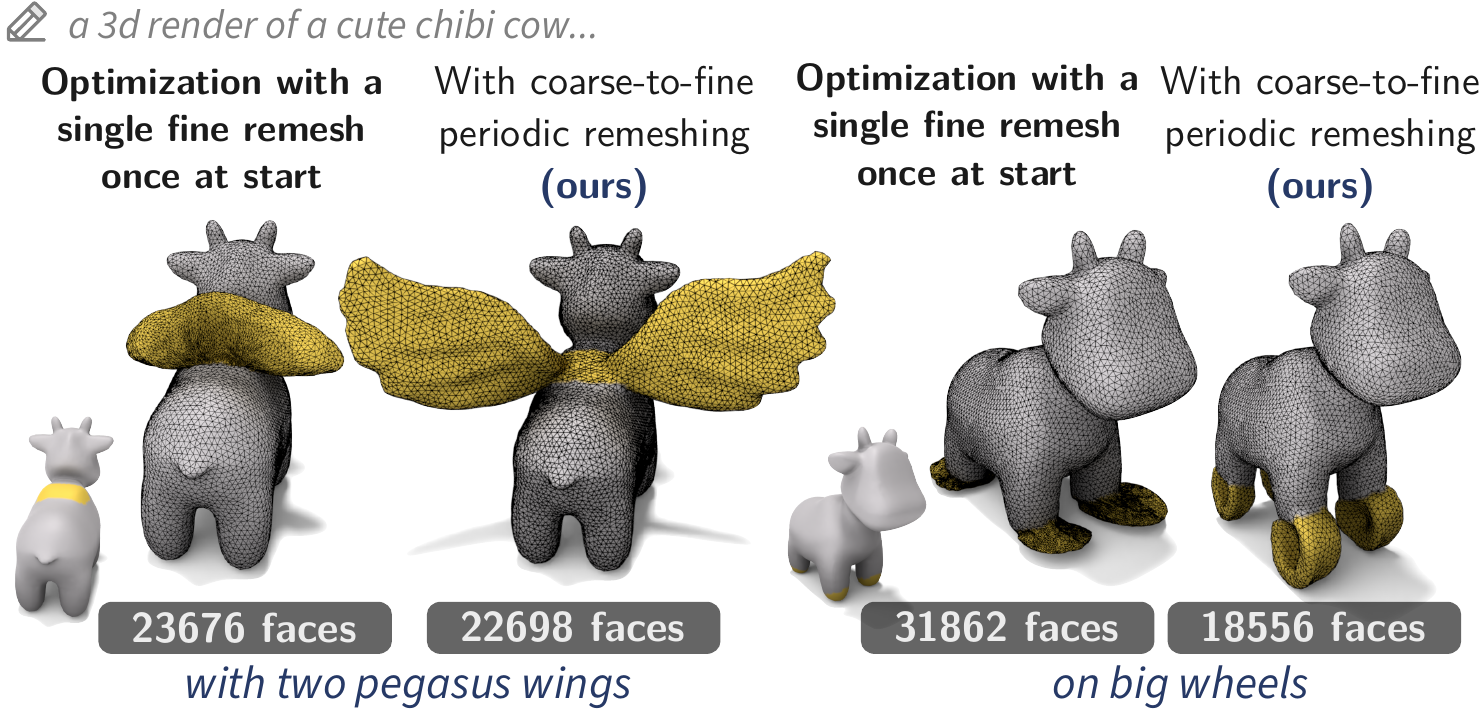}
        \caption{\centering\textbf{Ablation: no coarse-to-fine remeshing.}}
        \label{fig:remeshonceablation}    
    \end{subfigure}
    \begin{subfigure}[b]{0.3\linewidth}
        \includegraphics[width=0.99\linewidth]{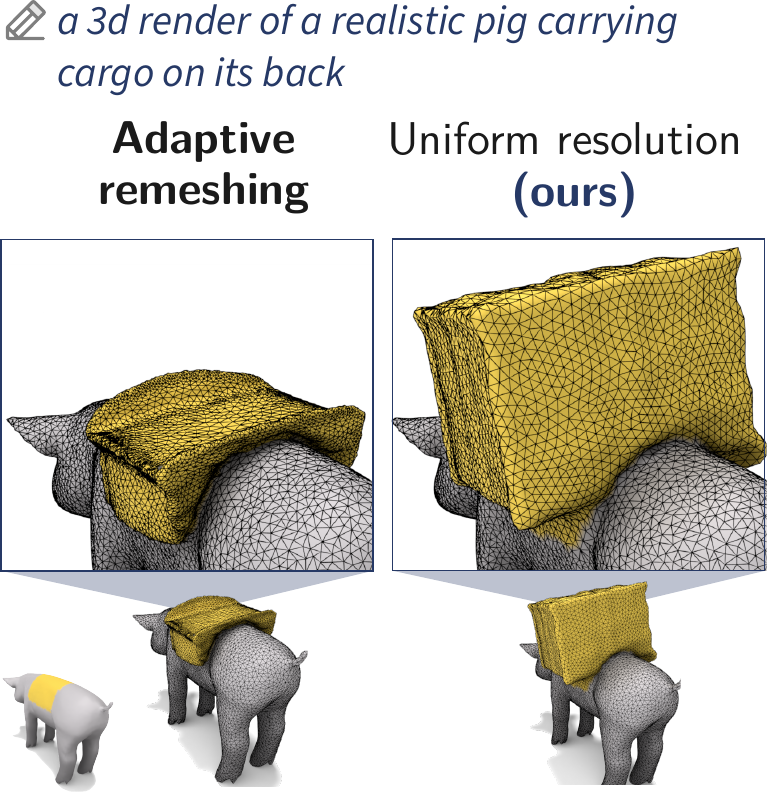}
        \caption{\centering\textbf{Ablation:\\adaptive remeshing.}}
        \label{fig:adaptivermshabl}
    \end{subfigure}
    \caption{\textbf{Ablations on remeshing.}
    \textbf{(a) No coarse-to-fine remeshing:} Simply finely remeshing the region \textit{only once at the start}, even to higher resolutions than the final resolution of our coarse-to-fine schedule, fails to produce quality growths. This shows the significance of a coarse-to-fine remeshing schedule.
    \textbf{(b) Adaptive remeshing:} Using adaptive remeshing (giving more resolution to high-curvature areas) rather than uniform isotropic remeshing also breaks the coarse-to-fine property, compromising the ability to grow large geometries.
    }
\end{figure}

Central to our method is the addition of the optimizable scale parameter, the optimizer state interpolation through the remeshing, and the coarse-to-fine remesh schedule. We validate these design choices with ablation experiments. 
In \cref{fig:noscaleabl}, we show that the scale parameter, $\qq_k^\mathrm{scale}$ (\cref{sec:deformqty}) is critical for growing large parts and appendages, with detail and sharpness even at extremities.
In \cref{fig:interpabl}, we show that when optimizer state interpolation is disabled, the deformation becomes unstable, as the optimization history is wiped after each remesh. This setting leads to two undesired effects: uneven results with geometric artifacts, as well as increased face counts. 
In \cref{fig:remeshonceablation}, we stress the importance of our coarse-to-fine remeshing schedule: 
remeshing to high resolution once complicates the optimization landscape and compromises expressivity. Similarly, \cref{fig:adaptivermshabl} shows that using adaptive remeshing \cite{dunyach2013adaptive} (\ie giving more resolution to higher-curvature regions), also fails to support large growths. Committing to dense triangles early, especially in areas undergoing rapid shape change whose curvature is not yet finalized, compromises the coarse-to-fine property. These results indicate that uniform, gradual remeshing allows better exploration of the deformation space, allowing efficient evolution of large and detailed geometry, as also visualized in \cref{fig:evolution}.

\subsection{Applications and Workflows}
\label{sec:workflows}
\begin{figure}[b]
    \centering
    \includegraphics[width=0.75\linewidth,trim=0 10 0 40]{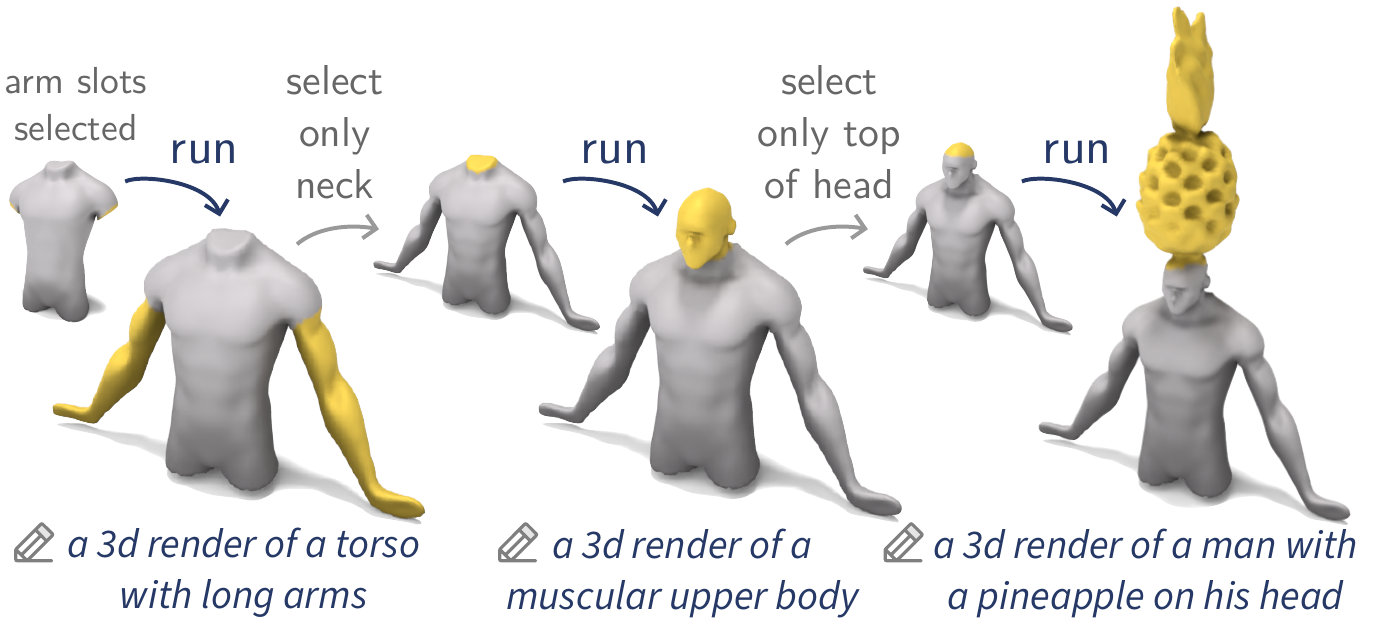}
    \caption{\textbf{Iterative editing.} We progressively add parts to the torso mesh by iterative selections (including on top of previously grown parts) and prompts.}
    \label{fig:torsoiterative}
\end{figure}

\begin{figure}[t]
    \centering
    \includegraphics[width=0.8\linewidth]{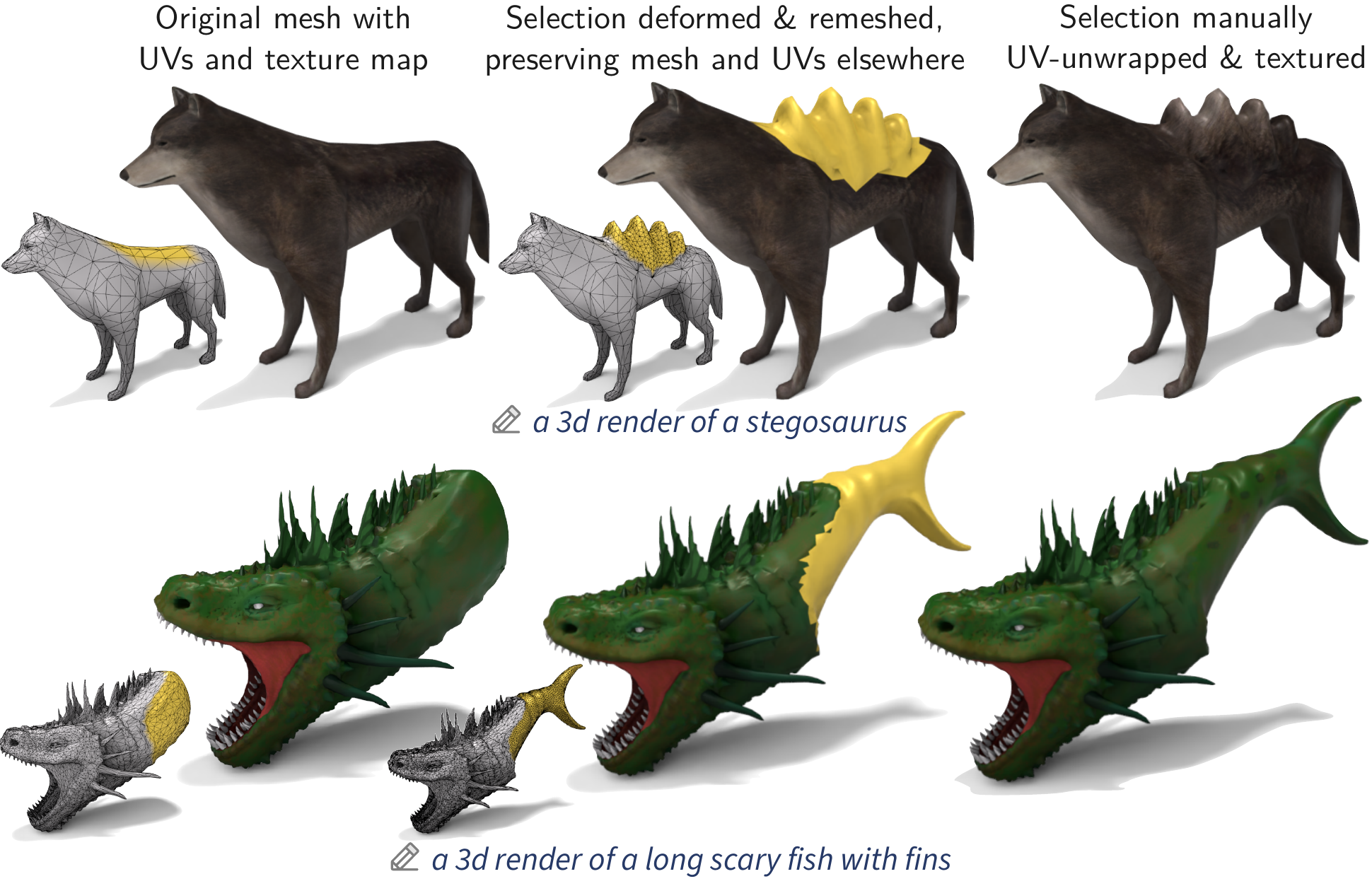}
    \caption{\textbf{Workflow on textured meshes.} Despite changing connectivity, our method interpolates vertex attributes inside, and fully preserves correspondence, geometry, and data of mesh elements outside the region selected for deforming and remeshing. This includes \eg corner UVs. We demonstrate the workflow of deforming and remeshing a region on a mesh with existing UVs and texture: the rest of the mesh is unchanged, and the user can choose to unwrap and texture the newly grown geometry.}
    \label{fig:uvtex}
\end{figure}

We demonstrate some extended workflows using our method. 
\cref{fig:torsoiterative} showcases a workflow comprising multiple prompts: we can iteratively add parts (arms, head, and a pineapple on the head) to the torso with high detail, including on top of previously grown parts.
\cref{fig:uvtex} shows a workflow on a textured mesh. Our method propagates vertex data within the selection region, and keeps all correspondence, corner UVs, and data outside the selection region. The new geometry can be easily re-unwrapped and retextured.

\subsection{Limitations} \label{sec:limitations}
Running time is dominated by the slow supervision, being SDS-based: on a single L40S GPU, a local growth run (2600 epochs) takes 85-90 minutes, and a global run (2200 epochs) 70-75 minutes. However, epochs earlier than this may already sufficiently satisfy the prompt.
The use of a Poisson solve also requires the mesh to be manifold, a restriction shared by all other such Poisson solve-based methods (Geometry in Style \cite{dinh2025geomstyle}, MeshUp \cite{kim2025meshup}, other NJF \cite{aigerman2022neural}-based systems.)
However, integrating non-manifold mesh Laplacians \cite{sharp2020nonmaniflaplacian} and non-manifold connectivity edit operations \cite{liu2025simplifying} is a promising direction for future work.

\section{Conclusion} \label{sec:conclusion}
We presented a technique for learning to deform and change the triangulation of a mesh using noisy visual losses.
Our framework optimizes a controllable, robust, smooth per-vertex deformation quantity instead of raw vertex positions. Our remesher propagates this quantity's optimization state across discrete remeshing operations, allowing optimization remain coherent despite the inherently non-differentiable nature of connectivity changes. As a result, our method smoothly molds meshes to have new extended parts and detail with stability, fidelity, triangle quality and efficiency.

In the future, we are interested in applying this framework for learning desirable triangulations and for quad meshing. In our current formulation, all the remeshing operations (edge collapse, splits, flips) are genus-preserving. Looking ahead, we aim to incorporate mesh tearing and merging operations that would allow the optimization to introduce topological holes and genus modifications.

\section*{Acknowledgments}

This work was supported by NSF grants \#2304481, \#2335493, \#2402894, \#2542757, BSF grant \#2022363, gifts from Adobe, Snap, and Google, and the Bennett Family AI + Science Collaborative Research Program. We thank the members of the 3DL lab for valuable discussions and feedback; we are grateful for the computing resources, services, and staff expertise at the University of Chicago.

\bibliographystyle{splncs04}
\bibliography{references}












\clearpage

\appendix

\setcounter{equation}{6}

\ifsigsubmit
    \setcounter{figure}{16}
    \begin{abstract}
    \end{abstract}
    \makeatletter
    \def\@title{Supplementary Material for \ourtitle}
    \def\@concepts{}
    \def\@keywords{}
    \@ACM@nonacmtrue
    \def\@mkbibcitation{}
    \makeatother
    \maketitle
\else
    \setcounter{figure}{12}
    \begin{center}
        \textbf{\Large\ourtitle}\\
        \vspace{0.2cm}
        {\Large Supplementary Material}
    \end{center}
\fi

\section{Hyperparameters}

\subsection{Optimization hyperparameters}
For local growths, we use a constant learning rate of 0.00175, and for global deformations, we use 0.00150. The Adam optimizer hyperparameters are $\beta_1 = 0.95$, $\beta_2 = 0.999$.

\def\degree{^\circ}

Each epoch uses 8 views sampled uniformly from an elevation range of $0$ to $60\degree$, distance range of $2.2$ to $2.8$, and a full $360\degree$ azimuth range. Local runs sample FOV uniformly from $55\degree$ to $80\degree$, while global runs use a constant $80\degree$ FOV. For differentiable rendering, we follow the use of \texttt{nvdiffrast} as in Hasselgren \etal~\cite{Hasselgren2021nvdiffmodeling}; for shading, we use a gray $(0.5,0.5,0.5)$ diffuse material and randomly placed point lights at a light power of 5.

The classifier-free guidance weight for CSD \cite{decatur2024paintbrushcsd} is $100$. The CSD method has weights for each diffusion stage; we use two stages of DeepFloyd IF. The weight for stage 1 diffusion is kept at 1.0; the weight of stage 2 diffusion is on a schedule. For all local growths, we use a schedule that linearly ramps from 0.0 to 0.25 for 1000 epochs, then from 0.25 to 0.4 for 700 epochs. For all global deformations, we use a schedule that goes from 0.0 to 0.2 for 1000 epochs, then 0.2 to 0.3 for 700 epochs.

Additionally, within one epoch, on the same batch of render images, we take CSD loss, backpropagate, and update the deformation quantity \emph{twice} before moving onto the next epoch. This technique empirically leads to sharper results and was also used in the official implementations of MeshUp \cite{kim2025meshup} and Geometry in Style \cite{dinh2025geomstyle}.

\section{More evaluation details}

\subsection{Quantitative evaluation shapes and prompts}
We use 64 shape-prompt pairs for the quantitative evaluations. We provide the result files for all the runs evaluated across all methods in the \texttt{evaluated-runs} folder in the supplementary material, alongside a Python script that visualizes the result shape, the selected region on both, as well as the prompt. The source shapes are saved and visualized in the run files in the \texttt{ours} directory; for a given prompt, all methods receive the same source shape after the same preprocessing that we do (\ie normal inflation, for a fair starting point for growths). We note that MagicClay SDF optimization crashes early in many runs (especially runs with all vertices selected), with only 49 out of 64 finishing with a result.

These data files are in the full supplementary data .zip, available on our \href{https://threedle.github.io/radmesh}{project page}.

\subsection{Instant3dit starting mask}
Because the Instant3dit \cite{barda2025instant3dit} inpainting method can only modify within a given mask geometry, its result (and the extent of its added/grown geometries) is heavily reliant on the shape of the mask. For the qualitative evaluation (Figure 9) we choose an early epoch from our method's runs to use as a mask for Instant3dit. The mask also receives a dilation of 0.25 (a hyperparameter exposed by Instant3dit).  Without this choice of mask, and using only the starting (inflated) selection, even with dilation, Instant3dit yields subpar results. However, as this selection process of a mask from our intermediate runs is highly ad-hoc, we opt not to use it for the quantitative evaluation, where we only give Instant3dit the selection mask (with dilation), rather than an intermediate epoch from our run.

\subsection{CLIP evaluation settings}
We sample 32 views for each result shape, using four elevations $(15\degree, 30\degree, 45\degree, 60\degree)$ and 8 evenly spaced azimuths from the full circle at each elevation. We use CLIP ViT-B/32 to evaluate the cosine similarity between the normalized embeddings of the images and of the text prompt.

\subsection{VQA score evaluation settings}
For the VQA score evaluation, we use 16 views for each result shape, with two elevations $(30\degree, 60\degree)$ and 8 evenly-spaced azimuths each elevation. We use the CLIP-FlanT5 VQA score.

\ifsigsubmit
\else
    
\fi

\section{Remeshing performance}
In the context of a run, the remeshing cost is negligible; a mesh with 18k faces takes about 0.2s to remesh w/ interpolation + 0.9s for Laplacian factorizing \& Procrustes matrix precomputations. (A mesh with 58k faces, 0.5s and 3.2s respectively.) Over a run of \eg 2600 epochs, 26 remeshes happen, taking $\sim$ $0.5\%$ of total time.

\section{More ablations}
\subsection{No leaky ReLU scale floor}
\begin{wrapfigure}[7]{r}{0.37\linewidth}
\vspace*{-3\intextsep}
\hspace*{0.6\columnsep}\includegraphics[width=0.9\linewidth,trim=0 1cm 0 0]{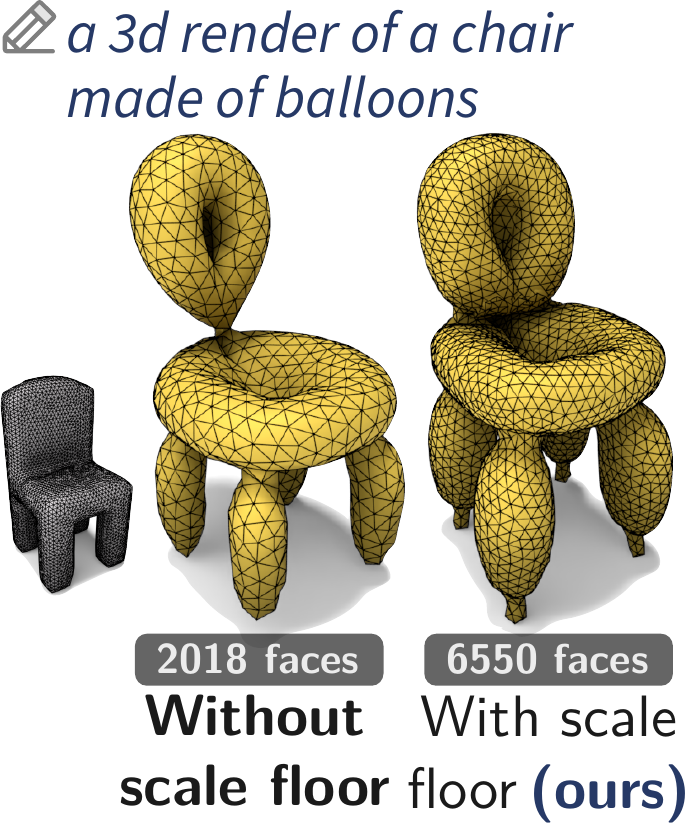}
\end{wrapfigure}
The 0.98 leaky ReLU floor for the scale parameter is meant to prevent a tendency towards excessive global shrinking that may cause loss of detail and volume. This does not hinder the ability for the optimization to form meaningful local shrinkages and indents (\eg tapered balloon lobes.)

\clearpage
\subsection{No initial inflation}
\begin{wrapfigure}{r}{0.4\linewidth}
\vspace*{-1.2\intextsep}
\includegraphics[width=1.02\linewidth,trim=0 1cm 0 0]{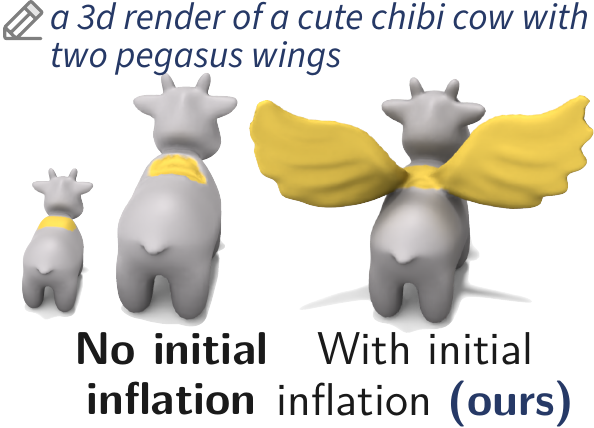}
\end{wrapfigure}
We show that initial inflation is necessary for some regions, such as the selection on the back of Spot, to begin deforming according to the prompt. We posit that for small, flat, or easily occluded regions, initial inflation gives a seed geometry that is larger in the renders, capturing SDS supervision signal on more surface area, thus kickstarting growth. Without it, some regions fail to grow.

\section{Deformation method details}

\paragraph{Procrustes solve}

The matrix representation for the Procrustes solve in equation (2) in the main paper is the following, using the notation from \citet{dinh2025geomstyle}.
At a vertex $k$, recall the neighborhood $\NN_k$ consisting of the ``spokes and rims'' halfedges: the halfedges radiating from vertex $k$, their twins, and the counterclockwise halfedges opposite $k$. Given the vertex's (face area-weighted) normal $\mathsf u_k$ on the \emph{current base mesh}, and the current deformation variable's direction component $\qq_k^\mathrm{dir}$ (an $\mathbb R^3$ vector, normalized to unit length before this computation), a matrix $\mathsf X_k$ is first computed:
\begin{align}
\mathsf X_k = 
\begin{bmatrix}\mathsf E_k & \uu_k\end{bmatrix} 
\begin{bmatrix}\mathsf W_k & \\ & \lambda a_k
\end{bmatrix}
\begin{bmatrix}\mathsf E_k^\top \\ {\qq_k^\mathrm{dir}}^\top \end{bmatrix}
\end{align}
where $\mathsf E_k$ is a $3 \times |\NN_k|$ matrix with the vectors of the halfedges in $\NN_k$ (in the current base mesh) as its columns, and $\mathsf W_k$ is a $|\NN_k| \times |\NN_k|$ diagonal matrix with the edges' cotangent weights (also of the current base mesh) on the diagonal. Then, taking the SVD of $\mathsf X_k$, the rotation matrix $\mathsf R_k$ is then found as:
\begin{align}
\mathbf U_k \mathbf \Sigma_k \mathbf V_k^\top &= \mathsf X_k \\
\Rr_k = \mathbf V_k \mathbf U_k^\top
\end{align}
with an appropriate multiplier on a column of $\mathbf U_k$ to ensure $\det(\Rr_k) = 1$.

\paragraph{Global solve}

Once the local step transform matrices $\Tt_k$ have been found (by the above Procrustes step and equations (3) and (4) in the main paper), equation (6) in the main paper can be solved \cite{liu2021normal} as the following matrix equation, a Poisson equation in $\VV'$:
\begin{equation}\label{eq:globalsolve-matrix}
L\VV' = 
\begin{bmatrix}
\operatorname{rhs}(1)^\top \\
\vdots\\
\operatorname{rhs}(|\VV|)^\top
\end{bmatrix}
\end{equation}
{\small\begin{equation}
\operatorname{rhs}(k)= \!\! \sum_{\scriptscriptstyle (k,m,n)\in \NN^F_k}{ \!\! \frac{\Tt_k + \Tt_m + \Tt_n}{3} \left(\frac{w_{km}}2\ee_{km} + \frac{w_{kn}}2 \ee_{kn} \right)}
\end{equation}}
\noindent where $L$ is the cotangent laplacian, $\VV$ is the current base mesh's vertex matrix (of shape $|\VV|\times 3$), $\NN_k^F$ is the set of faces adjacent to vertex $k$; each face has the vertices $(k,m,n)$ without loss of generality (\ie rotating vertex indices such that $k$ is first); $w_{km}, w_{kn} $ are the undirected cotangent weights of edges $(k,m), (k,n)$.

\paragraph{Restricting deformation to a selected region}

As described above, equation 6 in the main paper
becomes the Poisson equation \cref{eq:globalsolve-matrix}, which has form $LX=B$ with the cotangent Laplace operator as the system matrix. Solving a system with pinned vertices amounts to solving a modified system $L'X = B'$ where 
\begin{equation}
L' = \left[\text{$L_{jk}$} \mid j,k \in \{1\dots|\VV|\} \wedge \mathsf{sel}(j)=\mathsf{sel}(k)=1 \right]
\end{equation}
(\ie $L$ without the rows and columns corresponding to unselected vertices), and the adjusted right-hand side $B'$ defined as
\begin{align}
B'&=\left[\tilde B_{\text{row }k} \mid k \in \{1\dots|\VV|\} \wedge \mathsf{sel}(k)=1 \right] \\
\tilde B &= B - \tilde L \tilde V\\
\tilde L &= \left[
\text{$L_{\text{column }k}$} \mid k \in \{1\dots|\VV|\} \wedge \mathsf{sel}(k)=0 \}\right]\\
\tilde V &= [\pp_k \mid k \in \{1\dots|\VV|\}\wedge \mathsf{sel}(k)=0]^\top
\end{align}

\section{Generating with inverse rendering methods}

    \begin{figure}
        \centering
        \includegraphics[width=0.86\linewidth]{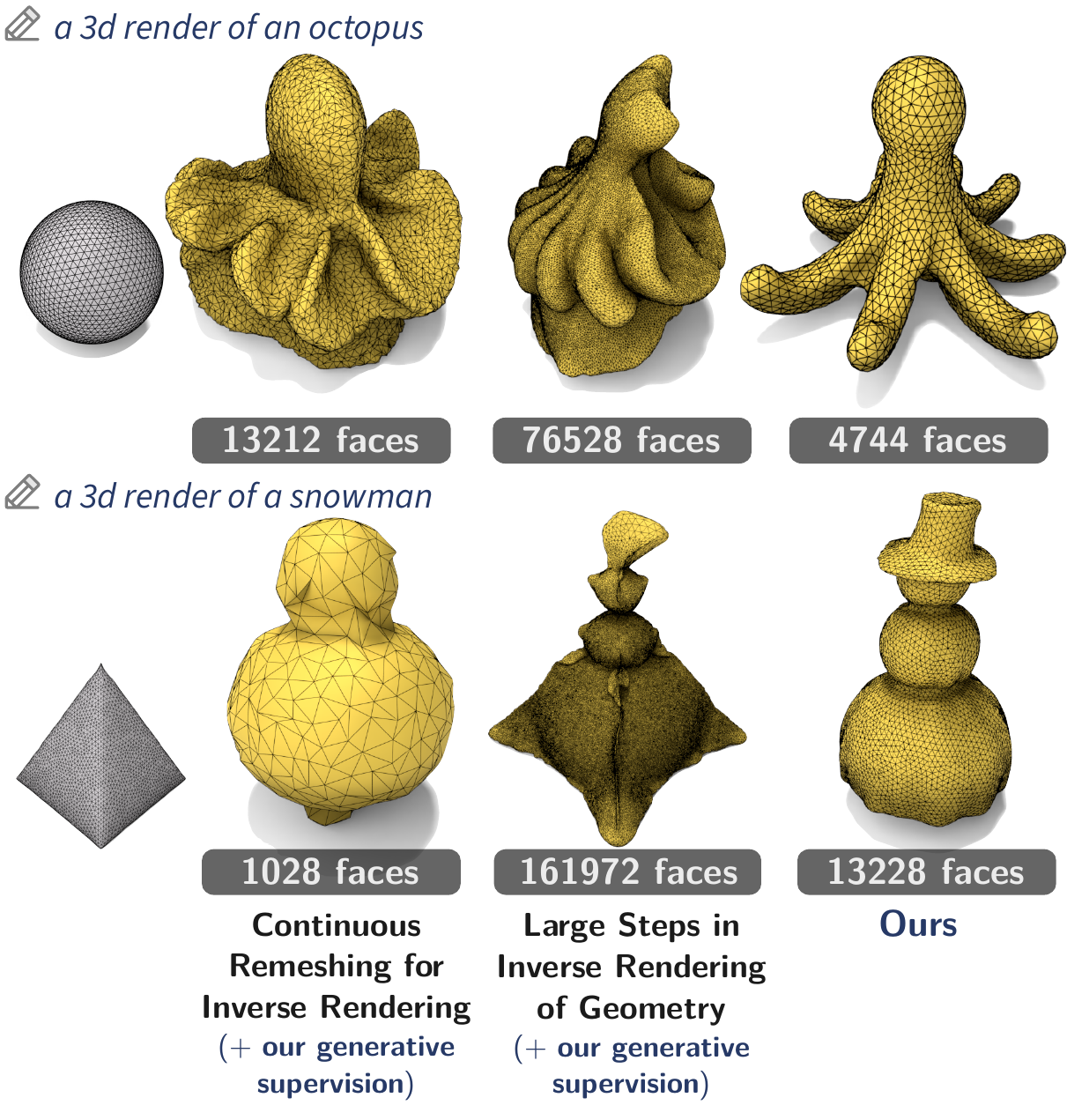}
        \caption{\textbf{Using inverse rendering methods with generative supervision.} We add our CSD-based generative semantic loss (replacing L1 supervision against ground truth target images) to the deformation-and-remeshing pipelines of the two methods \emph{Continuous Remeshing for Inverse Rendering} \cite{palfinger2022continuousremeshing} and \emph{Large Steps in Inverse Rendering of Geometry} \cite{nicolet2021largesteps}. Both methods, even granted some bespoke hyperparameter tuning to accommodate the new task, fail to achieve the large-scale geometric changes required to form not just high-level structure but also clear new parts \eg the octopus tentacles, the snowman's hat. Our method achieves these structures while maintaining smoothness, unlike the noisy surface elements of \emph{Continuous Remeshing}. Our method is also triangle-efficient, unlike the resolution induced by the remeshing requirements of \emph{Large Steps.}}
        \label{fig:contremeshlargesteps}
    \end{figure}

In \ifsigsubmit{Fig. 16 in the paper}\else\cref{fig:contremeshlargesteps}\fi, we demonstrate the suitability of our method (our proposed deformation quantity and our use of remeshing) for the generative task. We contrast this to other similar studies that propose their deformation and remeshing within the inverse rendering task. Namely, we compare to \emph{Continuous Remeshing for Inverse Rendering} \cite{palfinger2022continuousremeshing} and \emph{Large Steps in Inverse Rendering of Geometry} \cite{nicolet2021largesteps}. The former proposes direct vertex positions as the deformation optimization quantity and uses vertex velocity and momentum heuristics based on Adam to perform remesh operations. The latter proposes a Laplacian-coordinate-based deformation quantity that involves a global solve (somewhat similar to our method), and uses the \citet{botsch2004remesh} remesher with a coarse-to-fine schedule. Their schedule is sparser (i.e. only once every about 1000 epochs) with a fast resolution ramp-up (i.e. halving of average edge length upon each remesh).

Both methods in their intended use case employ direct L1 loss against target ground-truth renders as supervision. We replace this with the cascading score distillation-based (CSD) \cite{decatur2024paintbrushcsd} generative supervision which we use in our method as a semantic loss against a text prompt. We also afford each method with some hyperparameter tuning, given the very different task from what the original implementations assume. We run each method for 2600 epochs with our CSD and view batch sampling settings.

We observe that while the semantic loss, in all methods, leads to progress towards the desired shape specified by the prompts, only our method achieves clear formation of new geometric parts in addition to high-level structure, \eg octopus tentacles, or the snowman's hat. These growths require strong large-scale growing and shrinking (\eg to form the indents between tentacles; to form the edges of the top hat). \emph{Continuous Remeshing} suffers from noisy triangle elements. \emph{Large Steps} suffers from both asymmetric, unclean formation of global geometry  and excessive resolution in order to form detail, given its remeshing schedule that sets the target edge length to be half the average current edge length.

\section{Evolutions visualized}
Please see the accompanying \texttt{evolutions.mp4} in the supplementary material for some videos of the evolutions of shapes during optimization. 
The full supplementary data .zip can be found on our \href{https://threedle.github.io/radmesh}{project page}.


\end{document}